\documentclass[aps,pre,showpacs]{revtex4-2}
\usepackage{color}
\usepackage{amssymb}
\usepackage{amsmath}
\usepackage{bm}
\usepackage{graphicx}
\usepackage{overpic}
\usepackage{comment}
\usepackage{epstopdf}
\usepackage{tikz}

\newcommand{\ve}[1]{\ensuremath{\mbox{\boldmath$#1$}}}
\newcommand{\ma}[1]{\ensuremath{\mathbb{#1}}}

\DeclareMathOperator{\ku}{Ku}
\renewcommand{\Re}{{\mathrm Re}}
\newcommand{\Pe}{{\mathrm Pe}_{\mathrm r}}

\begin{document}
\title{Alignment of elongated swimmers in a laminar and turbulent Kolmogorov flow}

\author{M. Borgnino$^1$\footnote{Current affiliation: Department of Earth and Environmental Sciences, Universit\`a di Milano - Bicocca, Milan, Italy}}
\author{G. Boffetta$^1$}
\author{M. Cencini$^2$}
\author{F. De Lillo$^1$}
\author{K. Gustavsson$^3$}

\affiliation{$^1$Dipartimento di Fisica and INFN, Universit\`a di Torino, via P. Giuria 1, 10125 Torino, Italy}
\affiliation{$^2$Istituto dei Sistemi Complessi, CNR, via dei Taurini 19, 00185 Rome, Italy and  INFN, sez. Roma2 ``Tor  Vergata''}
\affiliation{$^3$Department of Physics, Gothenburg University, 41296 Gothenburg, Sweden}

\begin{abstract}
  Many aquatic microorganisms are able to swim. In natural environments
  they typically do so in the presence of flows. In recent years it
  has been shown that the interplay of swimming and flows can give
  rise to interesting and biologically relevant phenomena, such as
  accumulation of microorganisms in specific flow regions and local
  alignment with the flow properties. Here, we consider a mechanical
  model for elongated microswimmers in a Kolmogorov flow, a prototypic
  shear flow, both in steady and in turbulent conditions.  By means of
  direct numerical simulations, supported by analytical calculation in
  a simplified stochastic setting, we find that the alignment of the
  swimming direction with the local velocity is a general phenomenon.
  We also explore how the accumulation of microorganisms, typically
  observed in steady flows, is modified by the presence of unsteady
  fluctuations.
\end{abstract}


\maketitle

\section{Introduction}
\label{sec1}

Many microorganisms, from bacteria to microalgae, are motile and
capable of swimming \cite{guasto2012}. From biological fluids to lakes
and oceans, swimming microorganisms have to adapt their motility to
the surrounding fluid conditions in order to find food, escape from
predators, optimize light uptake and mate \cite{kiorboe2008}.  Complex
biological mechanisms are often involved in controlling the swimming
direction of microorganisms, by producing directed motility cued on
chemical or physical signals (such as chemo/photo/rheo taxis)
\cite{guasto2012}. The study of simple mechanistic models has shown
that the interplay of swimming and flow is relevant in determining the
dynamics of such organisms. The analysis of deterministic models for
ellipsoidal swimmers shows that, in a two-dimensional, laminar
Poiseuille flow the equations of motion for the swimmer are formally
akin to those of a pendulum. For a wide range of parameters,
the dynamics can be divided into two families of
trajectories~\cite{zottl2012nonlinear,zottl2013periodic}. Trajectories
close to the channel center encounter small velocity gradients and are
characterized by an oscillating swimming direction, analogous to
pendulum oscillations, while trajectories in the high-shear
regions, close to the walls, show tumbling motion, the equivalent of
librations. Recent experimental works have confirmed these
findings in a strain of smooth-swimming {\em E. Coli} in a Poiseuille
flow \cite{junot2019swimming}. Other parallel flows (such as the
Kolmogorov flow) show similar behaviours \cite{berman2021swimmer}. Such
swimming dynamics can cause accumulation (trapping) of elongated
microswimmers in high shear regions \cite{rusconi2014bacterial} (see
also \cite{bearon2015trapping}). Preferential trapping in high-shear
regions has been observed also for gyrotactic algae, although induced
by a different mechanism, i.e. the competition by flow-induced
rotation and upwards-biased swimming.
\cite{durham2009disruption,santamaria2014gyrotactic}.  The dynamics is
in general modified by the presence of noise
\cite{barry2015shear,berman2021swimmer, junot2019swimming} or steric
interactions with the walls \cite{zottl2012nonlinear}, which however
do not necessarily disrupt its effects.  In particular, in
Ref.~\cite{berman2021swimmer} it has been clarified that depending on
noise, swimming velocity and aspect ratio, swimmers may accumulate
  in regions of either high or low shear rate in the laminar
Kolmogorov flow. In general, it is now well recognized that the
interplay between swimming, flow and noise produces a variety of
effects in laminar, chaotic and turbulent flows, including
preferential accumulation of swimmers
\cite{durham2009disruption,santamaria2014gyrotactic,
  rusconi2014bacterial,barry2015shear,arguedas2020microswimmers,
  bearon2015trapping,dehkharghani2019bacterial}, clustering
\cite{zhan2014accumulation,pujara2018rotations}, enhanced transport
\cite{dehkharghani2019bacterial} and it can affect chemotaxis
\cite{rusconi2014bacterial} and biofilm
formation~\cite{yazdi2012bacterial}.

Besides accumulation, swimming
microorganisms have been observed to align with local flow properties. For
instance, even in presence of strong turbulent fluctuations,
elongated swimmers, such as bacteria, are found to align nematically
with vorticity \cite{zhan2014accumulation,pujara2018rotations}. We
have recently shown that elongated swimmers, swimming with constant
speed relative the surrounding flow, tend to align with the
local flow velocity in homogeneous and isotropic turbulence~\cite{borgnino2019alignment}.
This alignment can be interpreted as a
kinematic effect, i.e. it does not depend on the details of the flow
dynamics. Indeed, it has been observed in both direct numerical
simulations (DNS) and in kinematic models.
Breaking of fore-aft symmetry due to swimming results in correlations between the flow velocity and its gradients, which cause the swimmers to align with the fluctuating part of the flow.
It is however unclear whether this mechanism contribute to, or in principle compete with, the possible alignment with respect to a steady mean flow.

In this paper we investigate how the motility of microswimmers is
affected by a mean flow and the role played by turbulent fluctuations.
We use the Kolmogorov shear flow which, in the steady laminar form, is
a parallel flow with a periodic sinusoidal profile, thus
characterized by a series of channels with spatially alternating
flow direction. In turbulent conditions, the average velocity is still
periodic with superimposed fluctuations.  This configuration allows us
to study the effects of a mean flow on motility, ignoring the
complications introduced by the presence of walls.  We develop a
connection between the numerical results obtained with the
dynamical evolution of the Kolmogorov flow to a stochastic model for
turbulent fluctuations which is amenable to analytical investigation.
In the case of a steady Kolmogorov flow, the dynamical mechanism
leading to accumulation in specific flow regions has recently been
explained \cite{berman2021swimmer}. Our study widens the perspective
to include the effect of unsteady fluid dynamical fluctuations on both
accumulation and alignment.

The remaining of the paper is organized as follows. In Section
\ref{sec2} the equation for the model swimmers, the Kolmogorov flow
and the statistical model are introduced. In sections \ref{sec3} and
\ref{sec4}, the laminar and turbulent cases are discussed, followed by
the analytical and numerical analysis of the statistical model in
Sec.~\ref{sec5}. In the last section the results are summarized and
discussed. The appendices detail the derivation of the analytical
theory in two limits.
\section{Model equations}
\label{sec2}

\subsection{Kolmogorov flow}
In order to study the effects of a persistent flow, with non-vanishing
mean, on the orientation of elongated swimmers, we
consider the  incompressible velocity field $\bm u(\bm x, t)$
solution of the Navier-Stokes equations
\begin{equation}
\partial_t \bm u+\bm u\cdot \bm \nabla \bm u= -\bm \nabla p +\nu \Delta \bm u +\bm F\,,
\label{eq:1}
\end{equation}
with a sinusoidal forcing $\bm F=\left(F\cos(z/L),0,0\right)$.
The stationary solution is the laminar Kolmogorov flow
$\bm U(\bm x)=(U_0 cos(z/L),0,0)$ with $U_0=L^2 F/\nu$.
The Kolmogorov flow, introduced as a framework to investigate the
transition to turbulence, is probably the simplest shear
flow solution of the Navier-Stokes equation in the absence of
boundaries.

If $\Re=U_0L/\nu>\sqrt{2}$, the laminar solution becomes unstable to
perturbations at scales much larger than $L$ \cite{meshalkin1961investigation}
and, increasing $\Re$, the flow becomes chaotic and eventually turbulent.
It is remarkable, and very useful for theoretical and numerical studies, that
the mean profile of the resulting, statistically
stationary turbulent flow
retains the monochromatic shape of the laminar solution i.e.
$\langle \bm u(\bm x,t)\rangle=\bm U(\bm x)=(Ucos(z/L),0,0)$, where $\langle\cdots\rangle$ denotes an ensemble average with $U \le U_0$ due to turbulent drag.
As customary we decompose the flow as
\begin{equation}
\bm u(\bm x,t)=\bm U(\bm x) + \bm u'(\bm x,t)\,,
\label{eq:decomp}
\end{equation}
where $\bm u'$ are turbulent
fluctuations with vanishing average, $\langle \bm u'\rangle=0$.

Previous numerical investigations
\cite{musacchio2014turbulent} indicate that, for large
enough values of $\Re$, the root mean square (rms) value of the fluctuations of the $x$-component of the
velocity is proportional to the amplitude
of the mean profile, namely $u^\prime_{\rm rms}\simeq 0.54 U$.
In stationary conditions, the rate
$\langle \bm F \cdot \bm u \rangle = \frac{1}{2} F U$ at which kinetic energy is injected by the forcing
 equals the rate of dissipation
$\epsilon=\nu \langle {\rm tr}(\ma A^{\rm T}\ma A)\rangle$, where
$\mathbb{A}=\bm \nabla \bm u^{\rm T}$ denotes the velocity gradient tensor, so that $\epsilon=\frac{1}{2} F U$.
We recall that the dissipative length and time scales are defined as
$\eta=\nu^{3/4}/\epsilon^{1/4}$ and
$\tau_\eta=\sqrt{\nu/\epsilon}$.

\subsection{Stochastic Model}
\label{sec:SM}
To allow for an analytical treatment of the alignment dynamics,
we consider a stochastic flow
with a mean $\bm U(\bm x)$ given by the
Kolmogorov flow of given amplitude and turbulent fluctuations are replaced by a random, incompressible,
homogeneous and isotropic velocity field defined as
$\bm u'(\bm x,t)=\bm\nabla\wedge\bm B(\bm x,t)$, where the components of the vector potential $\bm B$ are taken to be
Gaussian random functions with zero mean and correlation
functions~\cite{Gustavsson2016}
\begin{equation}
\langle B_i(\bm x,t)B_j(\bm x',t')\rangle=\delta_{ij}\frac{\ell_{\rm f}^2 u_{\rm f}^2}{d(d-1)} \exp\Big(-\frac{|\bm x-\bm x'|^2}{2\ell_{\rm f}^2} - \frac{|t-t'|}{\tau_{\rm f}}\Big)\,.
\label{eq:2}
\end{equation}
Here $d$ is the spatial dimension. When $d=2$ we use the same
construction with $B_1=B_2=0$.  The stochastic velocity field $\bm
u'(\bm x,t)$ is characterized by its root-mean square amplitude
$u_{\rm f}$, a single length scale $\ell_{\rm f}$ and a single
Eulerian time scale $\tau_{\rm f}$. As a consequence of these
definitions, the correlation function (\ref{eq:2}) depends on a single
non-dimensional parameter, the Kubo number $\ku=u_{\rm f}\tau_{\rm
  f}/\ell_{\rm f}$, which measures the persistence of the flow.  In
the comparisons with turbulence, the Lagrangian time scale has been
found to be the relevant one \cite{Gustavsson2016}. Satisfactory
comparisons are obtained for $\ku=O(1-10)$. In the limit of large
values of $\ku$, the correlation function and hence the dynamics
becomes independent of $\ku$ when measured in units of $u_{\rm f}$ and
$\ell_{\rm f}$, with the possible exception of trapping of
trajectories occurring in the limit of frozen flow ($\ku \to \infty$).
The advantage of the stochastic model is the possibility to
investigate the effect of the time correlation of the flow on the
statistics of swimmers and, in the limit of small $\ku$, to obtain
analytical solutions by means of perturbation
theory~\cite{Gustavsson2016}.

It is useful to introduce time, velocity and length scales which can
be directly compared with the turbulent flow. For the stochastic model
we define $\tau_\eta=(\langle{\rm tr}(\ma A^{\rm T}\ma
A)\rangle)^{-1/2}=\ell_{\rm f}/(u_{\rm f}\sqrt{d+2})$, $u^\prime_{\rm
  rms}=u_{\rm f}/\sqrt{d}$, $u^\prime_{\rm rms}\tau_\eta=\ell_{\rm
  f}/\sqrt{(d+2)d}$. As a consequence,
$\ku=\tau_f/(\sqrt{d+2}\tau_\eta)$.  We remark that, within this
framework, the ratio $\alpha=U/u^\prime_{\rm rms}$ determining the
intensity of turbulent fluctuations and the dimensionless wave number
$\kappa=u^\prime_{\rm rms}\tau_\eta/L$ are independent parameters that
can take any value.  Given the definitions of $u^\prime_{\rm rms}$,
$\tau_\eta$ and $\alpha$, we can equivalently write $\kappa\sim
\ell_{\rm f}/L$ or $\kappa \sim \Re^{-1/2}/\alpha$.  On the contrary,
in the dynamic Kolmogorov flow, the values of $\alpha$ and $\kappa$
are given by the flow dynamics. Our DNS of the turbulent Kolmogorov
flow give $\alpha\simeq 2$~\cite{musacchio2014turbulent} and
$\kappa\sim 0.06$--$0.12$ for our range of Reynolds numbers.

\subsection{Model for elongated swimmers}
The swimmers are modelled as a dilute suspension (i.e. no feedback on the
fluid is considered) of small spheroidal, axisymmetric and neutrally
buoyant  particles moving with respect to the surrounding fluid with
constant speed $v_\mathrm{s}$~\cite{pedley87}.
The position $\bm x$ of a point-like, neutrally buoyant swimmer evolves
according to
\begin{equation}
\dot{\bm x} =
\bm u(\bm x,t) + v_\mathrm{s} \bm n\,,
\label{eq:3}
\end{equation}
where $\bm u(\bm x,t)$ is the fluid velocity at the particle
position.
The unit vector $\bm n$ determines the orientation of the
swimmer, and therefore its swimming direction, which evolves
according to  Jeffery's dynamics \cite{jeffery1922}
\begin{equation}
\dot{\bm n} = [\mathbb{O}(\bm x,t) + \Lambda \mathbb{S}(\bm x,t)]\bm n
-\Lambda [\bm n\cdot\mathbb{S}(\bm x,t)\bm n]\bm n\,,
\label{eq:4}
\end{equation}
where $\mathbb{O}$ and $\mathbb{S}$ are respectively the antisymmetric
(vorticity) and symmetric (strain) components of the velocity gradient matrix
$\mathbb{A}$.

The shape of the particle is quantified by the non-dimensional shape factor
$\Lambda=(a_\parallel^2-a_\perp^2)/(a_\parallel^2+a_\perp^2)$, where
$a_\parallel$ and $a_\perp$ are the particle sizes along and perpendicular
to $\bm n$, and is defined in the interval $[-1\!:\!1]$, with
$\Lambda\!=\!0$ corresponding to spheres,
$\Lambda=-1$ to flat disks and
$\Lambda=1$ to thin rods.
In this work we consider, for simplicity, and for its biological relevance, only the case of very elongated, rod-like swimmers.
As a consequence, in what follows we use $\Lambda\approx 1$.

The other parameter in the model is the swimming speed which will be
made dimensionless by normalizing with the typical velocity fluctuation
in the flow and therefore we define
the {\it swimming number} as $\Phi=v_\mathrm{s}/u^\prime_{\rm rms}$. The only exception to this definition will be the laminar case (see the next section), where there are no fluctuations and the swimming number will instead be defined with respect to the mean flow amplitude.

\section{Steady laminar Kolmogorov flow\label{sec3}}

In this section we briefly discuss the case of elongated swimmers in a
steady, laminar Kolmorogov flow $\bm U(\bm x)=U_0 \cos(z/L)$.
The equations ruling the position and orientation of the
swimmers read
\begin{eqnarray}
  \dot{x}&=& U_0\cos(z/L)+v_\mathrm{s} n_x  \label{eq:x1dot}\\
  \dot{y}&=& v_\mathrm{s} n_y \label{eq:y1dot}\\
  \dot{z}&=& v_\mathrm{s} n_z \label{eq:z1dot}\\
  \dot{n}_x&=& \frac{U_0}{L}\left[-(1+\Lambda)/2+\Lambda n_x^2\right] n_z \sin(z/L) \label{eq:nx1dot}\\
  \dot{n}_y&=& \frac{U_0}{L}\Lambda n_x n_y n_z \sin(z/L) \label{eq:ny1dot}\\
  \dot{n}_z&=& \frac{U_0}{L}\left[(1-\Lambda)/2+\Lambda n_z^2\right] n_x \sin(z/L)\,. \label{eq:nz1dot}
\end{eqnarray}

By inspecting the above equations it is clear that the relevant
dynamics involve only Eqs.~(\ref{eq:z1dot}), (\ref{eq:nx1dot}) and
(\ref{eq:nz1dot}) because these are independent of $x$, $y$ and
  $n_y$ governed by the remaining equations. Moreover, we notice
that if $n_y$ is initially zero it remains so. For
the full problem in three spatial dimensions, a treatment similar
to that discussed in Refs.~\cite{zottl2012nonlinear,zottl2013periodic}
can, in principle, be performed also here.  However, since the
steady  flow is not our primary interest in this study,
  we instead provide a simplified treatment with focus on the
  dynamics in two spatial dimensions (see also
Ref.~\cite{santamaria2014gyrotactic} for similar considerations).  We
also refer to a recent work \cite{berman2021swimmer} which presented a
detailed study of ellipsoidal swimmers, for arbitrary values of
  the shape factor $\Lambda$, in a two-dimensional laminar Kolmogorov
flow, subjected to rotational diffusion.

By expressing the swimming orientation as $\bm
n=(\cos\varphi,0,\sin\varphi)$ it follows that the three
equations (\ref{eq:z1dot}), (\ref{eq:nx1dot}) and (\ref{eq:nz1dot})
can be equivalently expressed as
\begin{eqnarray}
  \dot{z}&=&\Phi \sin\varphi \label{eq:zz}\\
  \dot{\varphi}&=& \frac{1}{2}\sin z(1-\Lambda\cos(2\varphi))\,, \label{eq:phi}
\end{eqnarray}
which are now written in non-dimensional variables by normalizing
lengths by $L$, velocities by $U_0$ and times by $L/U_0$; notice that,
unlike the rest of the paper, here $\Phi=v_\mathrm{s}/U_0$ as there are no
fluctuations, and that to ease the readability we use $z$ instead of $z'=z/L$ to denote the non-dimensional coordinate.

Combining the two equations above  yields the following equation (see Refs.~\cite{zottl2012nonlinear,zottl2013periodic} for a similar derivation in a Poiseuille flow and Ref.~\cite{berman2021swimmer})
\begin{equation}
\frac12 \sin z dz = \Phi \frac{\sin\varphi d\varphi}{1-\Lambda\cos(2\varphi)}, \label{eq:diff}
\end{equation}
which implies that, for $\Lambda>0$, the dynamics preserve the
following quantity

\begin{equation}
  \mathcal{H}  =-\frac{1}{2} \cos z
  +\Phi
  \frac{\mathrm{atanh}
    \left(\sqrt{\frac{2\Lambda}{1+\Lambda}}\cos\varphi\right)}
    {\sqrt{2\Lambda(1+\Lambda)}}
\,.
\label{eq:integral_motion}
\end{equation}

When $\Lambda=0$, spherical swimmers, the above dynamics become
Hamiltonian (with $\mathcal{H}=-1/2 \cos z + \Phi\cos\varphi$, i.e. the
Harper Hamiltonian), while for $\Lambda>0$ it is not
Hamiltonian. This difference has relevant consequences: if we start
many swimmers uniformly distributed in space and with uniformly drawn
orientations, i.e $(z,\varphi)$ uniformly distributed in
$[0:2\pi]\times [0:2\pi]$, for $\Lambda=0$ the uniform distribution is
conserved by the dynamics while for $\Lambda>0$ it is not. Therefore, the
swimmers may display some accumulation in specific regions of the
flow, as indeed observed e.g. in microfluidic shear flows
\cite{rusconi2014bacterial} (see also Ref.~\cite{bearon2015trapping}).

\begin{figure}[t!]
\centering
\includegraphics[width=1\textwidth]{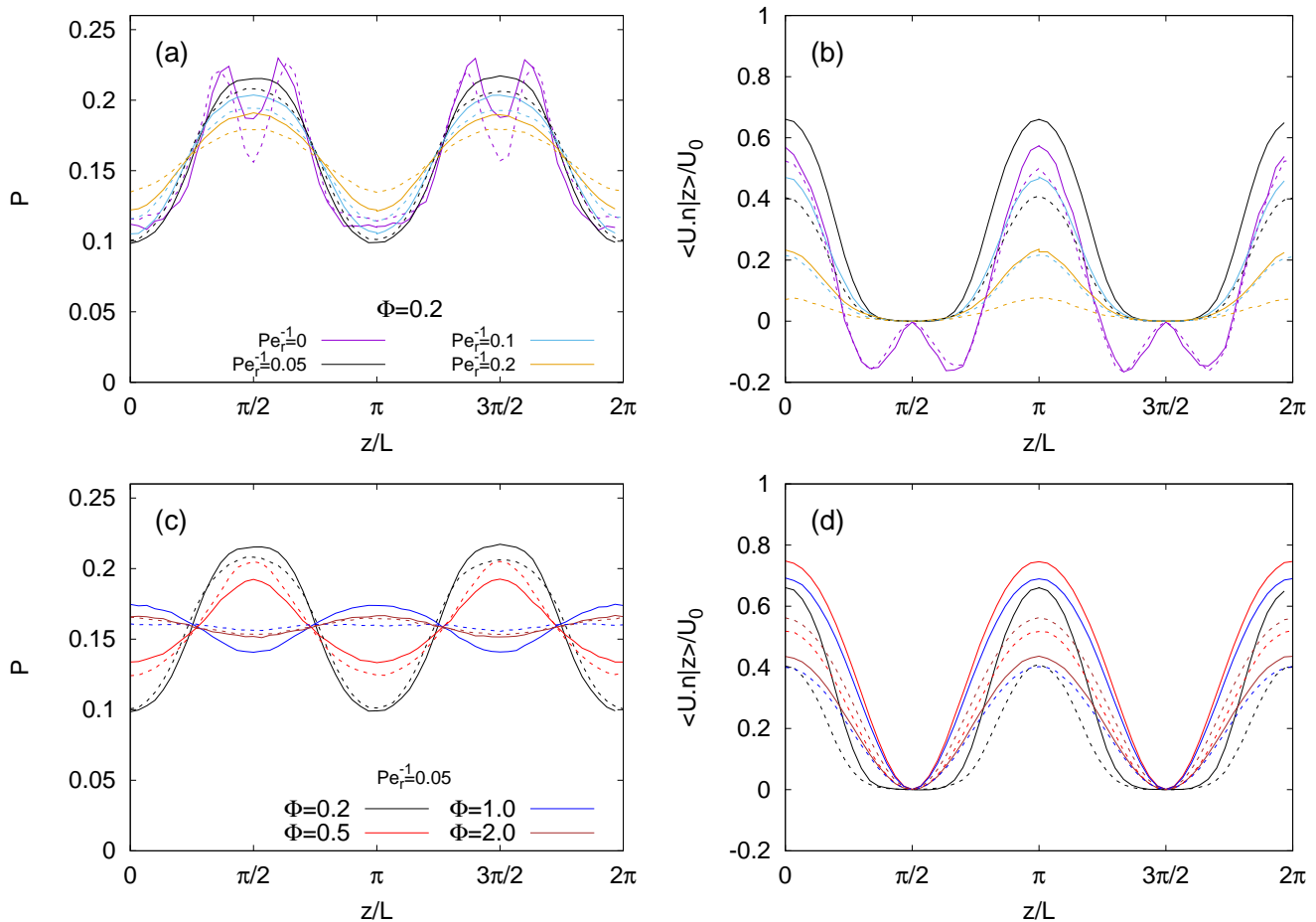}
\caption{(Color online) Results for the steady Kolmogorov
    flow. Microswimmer transverse PDF $P(z)$ and alignment in the
    steady Kolmogorov flow: (a,c) $P(z)$ vs $z/L$, (b,d) $\langle \bm
    U\cdot \bm n|z\rangle/U_0$ vs $z/L$ for $\Lambda=0.98$. In (a,b)
    we keep $\Phi=0.2$ fixed and vary the rotational noise,
    i.e. $\Pe$, as labelled; in (c,d) we keep the rotational
    noise fixed to $\Pe^{-1}=0.05$ and vary the swimming speed $\Phi$.
    Solid lines refer to the two-dimensional model, dashed lines to the full
three-dimensional one.}\label{fig:1}
\end{figure}

Recently, by considering turbulent and stochastic flows, we showed
\cite{borgnino2019alignment} that elongated microswimmers
($\Lambda>0$) tend to align their swimming direction to that of the
underlying advecting velocity field, a phenomenon which finds its
roots in the breaking of the fore-aft symmetry induced by swimming. It
is thus natural to ask whether such an alignment is present also in
the steady Kolmogorov flow. In order to explore such
alignment, we study  $\langle \ve n \cdot \ve U(\ve x)|z\rangle=\langle \cos\varphi \cos z|z\rangle$.
We solve Eq.~(\ref{eq:integral_motion}), with $\mathcal{H}$ replaced by its value ar $t=0$, for $\cos\varphi(t)$ to obtain
\begin{equation}
  \cos\varphi(t)\cos z(t)= \sqrt{\frac{\Lambda+1}{2\Lambda}} \tanh\left[
    \mathrm{atanh}\left(\sqrt{\frac{2\Lambda}{\Lambda+1}} \cos\varphi_0\right) +\frac{\sqrt{2\Lambda(1+\Lambda)}}{2\Phi}(\cos z(t)-\cos z_0)\right] \cos z(t)\,.
\label{eq:unDet}
\end{equation}
Averaging this equation numerically at a fixed value of $z$ over a
uniform distribution of initial coordinates $(z_0,\varphi_0)$, does
indeed give a positive correlation between swimming direction and
velocity. The alignment obtained agrees qualitatively with the
alignment observed in Fig.~\ref{fig:1}(b,d), although the simplified
description cannot reproduce the negative correlations of the
deterministic case. This difference is due to the
  fact that the uniform distribution we used to average the initial coordinates is not
  the stationary distribution of the model (see
  e.g. Ref.~\cite{berman2021swimmer} and Fig.~\ref{fig:2}b below). It
  is worth noticing that Eq.~(\ref{eq:unDet}) is valid for any
  $\Phi>0$, but does not apply when $\Phi=0$, where the $z$ and
  $\varphi$ dynamics decouple, thus the limit $\Phi\to 0$ is singular.
  The periodic dynamics on the iso-contours (\ref{eq:integral_motion})
  when $\Phi>0$ is fundamentally different from the periodic orbits
  (Jeffery orbits \cite{jeffery1922}) obtained when $\Phi=0$.  For the
  latter case the angular dynamics undergoes periodic motion with
  angular velocity $\Omega_0=\sqrt{1-\Lambda^2}\sin z_0/2$, driven by
  the mean flow shear, $\tfrac{\partial U_x}{\partial z}=-\sin z_0$,
  at the constant value of the transverse coordinate $z_0$.  The swimmer on
  the other hand, undergoes periodic dynamics in the $z$-$\varphi$
  space. Even though the angular dynamics of the swimmer is driven by
  the mean flow shear, the instantaneous angle given by
  Eq.~(\ref{eq:integral_motion}) is determined by the primitive
  function of the fluid gradient, i.e. the orientation is determined
  by the mean flow $\propto\cos(z)$ at the instantaneous position,
  Eq.~(\ref{eq:unDet}), rather than the mean flow gradient matrix.
  This shows that the periodic orbits of the swimmer are of a
  different nature than Jeffery's orbits.

We illustrate the phenomenology described above in Fig.~\ref{fig:1},
where we show (a,c) the probability density function (PDF) of swimmer
transverse position, $P(z)$, and (b,d) the conditional average $\langle
n_x \cos z|z\rangle$ for $\Lambda=0.98$; the reason for
choosing $\Lambda<1$ is the presence of a bifurcation point for $\Lambda=1$ at
which Eq.~(\ref{eq:phi}) possesses marginally stable fixed points, which are of
limited physical interest, corresponding to the swimmers oriented along the
flow. This behaviour is also reflected in the singularity in
Eq.~(\ref{eq:diff}).  We show the results both for the original
deterministic dynamics (\ref{eq:zz}-\ref{eq:phi}) and in the presence
of rotational noise that is added to Eq.~(\ref{eq:phi}) in the form of
a stochastic term, $\sqrt{2\Pe^{-1}} \eta(t)$, where $\Pe=U_0/(LD_{\rm
  r})$ denotes the rotational P\'eclet number, defined in terms of the
rotational diffusion coefficient, $D_{\rm r}$. We remark that the
results of simulations of the two-dimensional dynamics
(\ref{eq:zz}-\ref{eq:phi}) (solid lines) are qualitatively the same as
results of the original dynamics (\ref{eq:x1dot}-\ref{eq:nz1dot}) in
three spatial dimensions (dashed lines).

Figure~\ref{fig:1}(a) shows that swimmers tend to accumulate in
high-shear regions, as also observed in
Ref.~\cite{rusconi2014bacterial} for Poiseuille flows. As explained in
Ref.~\cite{rusconi2014bacterial}, this phenomenon is physically
induced by the fact that the swimmers rotate rapidly, causing them to
tumble in small loops in regions of intense shear. Rotational noise
can help stabilize this effect
\cite{rusconi2014bacterial,berman2021swimmer}, as can be seen in
Fig.~\ref{fig:1}(a), where the regions of depletion around $z/L\approx
\pi/2$ and $3\pi/2$ disappear in favour of a smooth maximum when a
small diffusivity is present. The effect is weakened for larger
rotational noise. However, as shown in Fig.~\ref{fig:1}(c), when the
swimming speed is increased, the swimmers instead accumulate in
regions with low shear and high flow velocity. The transition between
these two regimes is actually quite complex and a detailed study of
the dependence on rotational noise, swimming speed and swimmer shape
parameter $\Lambda$ is beyond the scope of this work. A complete
analysis can be found in Ref.~\cite{berman2021swimmer}, with a
detailed study of how the topology of phase space affects the swimmer
dynamics in the deterministic case and, for the case with noise, a
computation of the joint PDF of $z$ and $\varphi$.

Summarizing the above observations, in a steady Kolmogorov flow we
expect swimmers to align with the flow velocity for generic swimming
numbers and we expect some form of accumulation (i.e. non-uniform
spatial distribution) of the microswimmers. Both effects are
robust to the addition of rotational noise to the dynamics. Actually,
as shown in Ref.~\cite{berman2021swimmer}, a weak
noise has a stabilizing role on the accumulation. This is confirmed
by Fig.~\ref{fig:1}(a), while Fig.~\ref{fig:1}(b) shows the same
effect also on the alignment.

\section{Turbulent Kolmogorov flow: numerical simulations}
\label{sec4}

In this Section we study the effect of turbulent fluctuations on
accumulation and alignment of microswimmers.  Turbulence is a multiscale process, with fluctuations at many length
scales, characteristic times and intensity between the large scale
defined by the mean flow and the dissipative one, $\eta$
\cite{frisch1995turbulence}. This is of course very different from the
case of a mean flow with rotational diffusion, where there usually is a clear scale
separation between the deterministic component of
the flow and the short-time-correlated noise associated with the
diffusive behaviour.  In order to study the effects of turbulent
fluctuations on orientation and spatial distribution of swimmers, we
performed DNS of the Navier-Stokes equations
(\ref{eq:1}) with Kolmogorov forcing by means of a $2/3$ dealiased,
fully parallelized pseudo-spectral code in a cubic, triple-periodic
domain for three Reynolds numbers, namely $\Re=230, 730$ and $1350$.
The grid resolution $N$ was chosen as a function of the Kolmogorov scale $\eta$ (see Table~\ref{table1}) to guarantee that $k_{\rm max}\eta \gtrsim 1.23$ for all Reynolds
numbers, where $k_{\rm max}=N/3$ is the
maximum represented wave-number after de-aliasing.  At each
time step the values of the velocity field and its gradients were
interpolated onto particle positions to integrate the swimmer
dynamics (\ref{eq:3}-\ref{eq:4}).  Both fluid and swimmer equations
were advanced with a second-order Runge-Kutta scheme.

\begin{table}[h!]
\begin{tabular}{c|c|c|c|c|c|c}
$N$ & $\Re$ & $U$ & $u^\prime_{\rm rms}$ & $\epsilon$ & $\eta$ & $\tau_{\eta}$ \\ \hline
$128$ & $230$ & $0.23$ & $0.12$ & $9.31 \times 10^{-4}$ & $3.22 \times 10^{-2}$ & $1.04$ \\
$256$ & $730$ & $0.73$ & $0.38$ & $2.30 \times 10^{-2}$ & $1.44 \times 10^{-2}$ & $0.21$ \\
$512$ & $1350$ & $1.35$ & $0.76$ & $1.73 \times 10^{-1}$ & $8.72 \times 10^{-3}$ & $0.076$
\end{tabular}
\caption{Parameters used in the DNS of the turbulent Kolmogorov flow.
$N$ resolution, $U$ amplitude of the mean profile,
$\Re=U L/\nu$, $u^\prime_{\rm rms}$ rms of the fluctuation of the $x$
component of the velocity, $\epsilon=\nu \langle{\rm Tr}(\ma A^{\rm T}\ma A)\rangle$
mean energy dissipation, $\eta=(\nu^3/\epsilon)^{1/4}$ Kolmogorov scale,
$\tau_{\eta}=(\nu/\epsilon)^{1/2}$ Kolmogorov time scale.
The integral scale $L=1$ and the viscosity $\nu=10^{-3}$ are
fixed for all simulations.}
\label{table1}
\end{table}

\begin{figure}[h!]
\centering
\includegraphics[width=1.0\columnwidth]{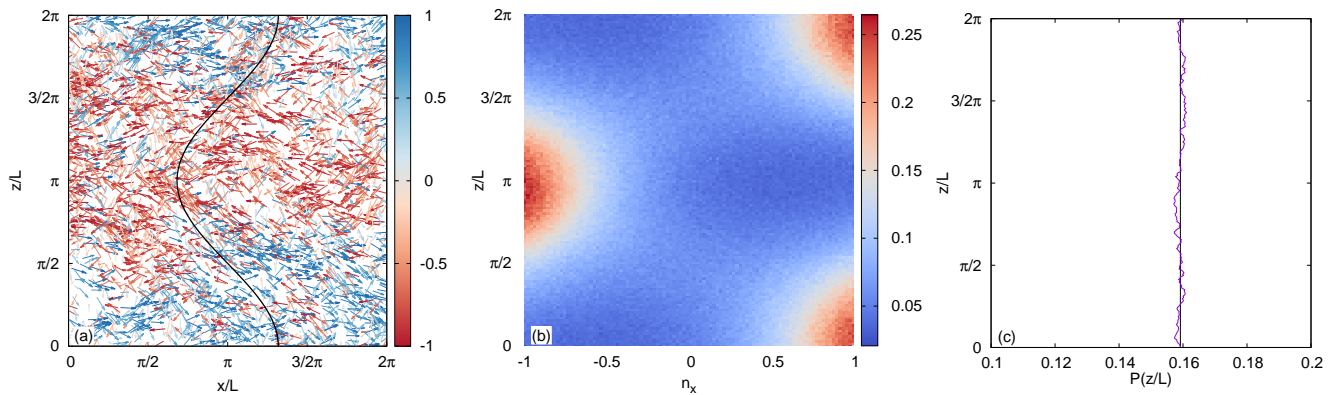}
\caption{(Color online) Results in the turbulent Kolmogorov flow: (a) swimmers' orientations in a turbulent
  simulation at $\Re=730$.  The swimmers are taken from a slab of
  thickness $0.4 L$ parallel to the mean shear plane. The arrows start
  from the particle positions and represent the projection of the
  swimming direction ${\bf n}$ onto the $xz$-plane.  The color map
  codes the intensity of $n_x$, i.e. the $x$-component of ${\ve n}$. The
  black line displays the mean velocity profile $U\cos(z)$.  (b): time averaged joint PDF $P(z,n_x)$ of swimmer
  transverse position ($z$) and $n_x$ for all the swimmers in the
  simulation domain.  Peaks are due to the alignment of the swimming
  direction with the mean flow profile.  (c) PDF of swimmer transverse positions
$P(z)$ obtained from $P(z,n_x)$ integrated over all
  possible $n_x$. Data refer to swimming number $\Phi=1.09$.}
\label{fig:2}
\end{figure}

In Fig.~\ref{fig:2}(a)  we plot the instantaneous
orientation of the swimmers in a slab of thickness $0.4~L$ in the
$xz$-plane.  The arrows, representing the projection of ${\bf n}$
onto the $xz$-plane, clearly indicate a strong alignment of the
instantaneous swimmer orientation with the mean velocity profile,
represented by the black solid line, proportional to $\cos z$.  In
order to better quantify the strength of this alignment, Fig.~\ref{fig:2}(b) shows the joint PDF $P(z,n_x)$ of the swimmer
position ($z$ coordinate) and swimming direction ($n_x$ component). A
strong correlation is evident with two peaks at $(z/L,n_x)=(0,1)$ and
$(\pi,-1)$, corresponding to the blue and red arrows in
Fig.~\ref{fig:2}(a), respectively.  In spite of the strong correlation
between mean velocity and swimming orientation, we do not observe
significant preferential accumulation,
i.e. swimmers are almost uniformly distributed along the inhomogeneous
direction $z$. This is shown in Fig.~\ref{fig:2}(c), where we plot the
distribution of $z$-swimmer positions $P(z)$ obtained by integrating
$P(z,n_x)$ over $n_x$. The same phenomenology is observed also for other values
of the parameters. This is consistent also with previous
results with other orientation mechanisms, such as gyrotaxis, where
accumulation in the Kolmogorov flow could be obtained only by artificially
reducing the intensity of turbulent fluctuations
\cite{santamaria2014gyrotactic}. We can therefore conclude that preferential
alignment is
an independent and more robust feature than preferential concentration for
microsmimmers in the Kolmogorov flow,
since the former is observed in both the laminar and in the turbulent case,
while the latter disappears in the turbulent flow.
We remark that we do not consider here possible small scale
spatial clustering (which is observed for swimming  \cite{zhan2014accumulation}
and passivley sedimenting \cite{ardekani2017sedimentation} elongated
particles in turbulence) but only inhomogeneities in
the distribution
on the scale of the mean flow.

\begin{figure}[h]
\centerline{\includegraphics[width=1.0\columnwidth]{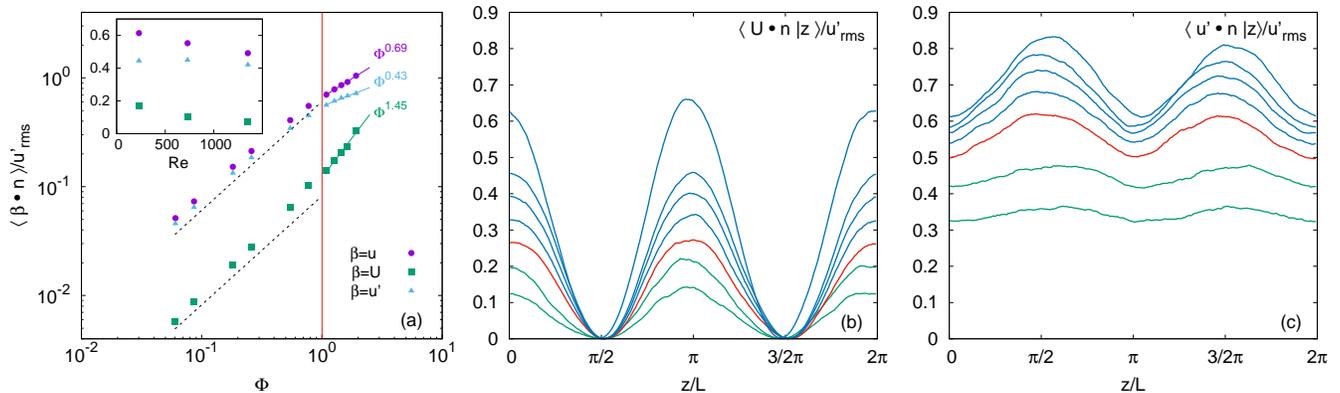}}
\caption{(Color online) Swimmer alignment in the
    turbulent Kolmogorov flow. (a) The scalar product $\langle \bm
  \beta \cdot \bm n \rangle$ averaged over space and time, as a
  function of the swimming number $\Phi$ with $\bm \beta=\bm u$
  (purple circles), $\bm \beta=\bm U$ (green squares)
  and $\bm \beta=\bm u'$ (blue triangles).
The indicated power-law behaviors are obtained from a fit of
the points at $\Phi>1$.
In the inset, the
  same quantities are plotted for fixed $\Phi\approx0.75$ for different
  $\Re$. (b) and (c) Modulation of the correlation between mean fluid
  velocity (b) or fluctuations (c) and swimmer orientation averaged
  over $x$, $y$ and time for different swimming numbers. From bottom to top
   $\Phi=0.54, 0.78$ (green
  curves), $\Phi=1.09$ (red curve), $\Phi=1.27, 1.45, 1.63, 1.92$
  (blue curves).  Simulation at $\Re=730$.  }
\label{fig:3}
\end{figure}

As in the previous section, we evaluate the alignment with the mean
flow by computing the scalar product $\langle \bm U (\bm x) \cdot {\bf
  n}\rangle$ of the swimming vector with the mean fluid velocity,
averaged over space and time. However, previous work showed that
swimmers can align with the local velocity even in homogeneous
turbulence \cite{borgnino2019alignment}. Such local alignment
could in principle disturb the alignment with the
mean flow.  Following the decomposition of the fluid velocity,
Eq.~(\ref{eq:decomp}), we therefore also consider the alignment with the
turbulent fluctuations $\langle \bm u^\prime (\bm x, t) \cdot {\bf
n}\rangle$ and with the complete fluid velocity $\langle \bm u (\bm
x, t) \cdot {\bf n}\rangle$, which is simply the
sum of the alignment with the average and fluctuating component (cf. Eq.~(\ref{eq:decomp}))
\begin{equation}
\langle \bm u \cdot \bm n \rangle=\langle \bm U \cdot \bm n \rangle+\langle \bm u' \cdot \bm n \rangle\,.\label{eq:decomp2}
\end{equation}

Figure~\ref{fig:3}(a) summarizes the
statistics of the alignment as a function of the swimming
number $\Phi$ for the simulation with $\Re=730$. We find that for
small values of $\Phi$, the alignment with the mean flow increases
linearly with the swimming number. The same behaviour is observed for
the alignment with the fluctuations and, as a consequence, with the
complete fluid velocity.  An alignment proportional to $\Phi$ was
already found for an isotropic turbulent flow, in the absence of a
mean flow \cite{borgnino2019alignment}. For larger values of $\Phi$,
the alignment with the mean flow increases faster, while that with the
full velocity (and fluctuations) is sub-linear.  The transition occurs at
$\Phi \simeq 1$ (vertical line in Fig.~\ref{fig:3}).
An explanation of the transition is that, for $\Phi>1$, $v_\mathrm{s}$ becomes
faster than the turbulent velocity, so that flow correlations along swimmer trajectories are dominated by swimming instead of transport by the fluid.
As a consequence, the Lagrangian correlation time of
fluctuations at scale $\ell$ seen by a swimmer can be estimated as
$\ell/v_\mathrm{s}$, so the orientation quickly decouples from the small-scale
fluctuations, while it remains correlated with the large-scale
gradients of the mean flow.
In other words, $\langle \bm U \cdot \bm n \rangle$ increases faster at
large $\Phi$ because the chaotic reorientation induced by flow velocity
fluctuations becomes less efficient, i.e. they average out.

Panels (b) and (c) in Fig.~\ref{fig:3} show the $z$-dependence of alignment through the conditional averages
$\langle \bm U \cdot \bm n|z \rangle$
and $\langle \bm u^\prime \cdot \bm n|z \rangle$ for different values of $\Phi$.
In all cases, we observe that $\bm n$ is always positively correlated with both the average and fluctuations of the
velocity field.
Moreover, the correlation increases with $\Phi$, in agreement with the results shown in Fig.~\ref{fig:3}(a).
Since the amplitudes of both the mean flow and
fluctuations are modulated along $z$, the profiles observed in
Fig.~\ref{fig:3}(b) and (c) are a combined result of the space dependence of
the velocity field and of modulations in the alignment of the swimmers.
In the case of the mean flow, Fig.~\ref{fig:3}(b), the maximal alignment
is observed where the flow is large and the gradients are small, i.e. around $z=0$ and $z=\pi$, while in the case of turbulent fluctuations,
Fig.~\ref{fig:3}(c), the alignment is stronger where the mean flow
is small, i.e. around $z=\pi/2$ and $z=3\pi/2$.

In order to investigate the dependence of the alignment on the Reynolds
number, we show the alignment statistics in the inset of Fig.~\ref{fig:3}(a)
with fixed $\Phi \approx 0.75$ for our three values of $\Re$.
We observe  that the correlation with the mean velocity
decreases with $\Re$ when rescaled with the typical velocity $u^\prime_{\rm rms}$
(which is proportional to $U$ as shown in Table~\ref{table1}),
while the correlation with the velocity fluctuations is almost independent
on $\Re$. This is remarkable since the ratio $U/u^\prime_{\rm rms}$ is independent
on $\Re$ (see Table~\ref{table1}). We comment further on this in the next section.

We remark that in order to have a constant swimming number $\Phi$ for
different values of $\Re$, the simulations were performed with
different swimming speeds, $v_\mathrm{s}=\Phi u^\prime_{\rm rms}$,
which increases with $\Re$ (see Table~\ref{table1}). Of course, if
interested in describing the effects of turbulence on a specific
motile species, we should consider a fixed $v_\mathrm{s}$ and we would
find that the alignment decreases with $\Re$ (not shown). This is in
agreement with what has been observed in an isotropic turbulent flow
\cite{borgnino2019alignment} and it is due to the weaker correlation
between velocity gradients, which orient the swimming direction, and
the velocity field.

\section{Unsteady Kolmogorov flow: stochastic model}
\label{sec5}

We now move to the statistical model introduced in
  Sec.~\ref{sec:SM}, for which we can choose the relative importance
of the Eulerian and Lagrangian time scales of the flow by tuning the
Kubo number.  For small values of $\ku$, the flow approaches the
laminar Kolmogorov flow discussed in Sec.~\ref{sec3} with a small
stochastic noise characterized by the Eulerian time scale $\tau_{\rm
  f}$. While, in the limit of large $\ku$, the
  disturbances become persistent. In this limit the relevant time scale
  is therefore Lagrangian in nature, because the statistics
  decorrelates due to transport by the mean flow or through swimming.
  To compare to our DNS of turbulence, we found, in agreement with previous studies
  \cite{Gustavsson2016,borgnino2019alignment}, that the Kubo number
  should be $O(10)$ (larger values give the same results). Further, we match both the Kolmogorov time scale
  and Taylor length scale in the two flows.  Moreover, we choose the
  parameters $\alpha=1.92$ and $\kappa=0.08$ according to our DNS with
  $\Re=730$ in Table~\ref{table1}.  These choices are expected to
  provide, at least, a qualitative agreement between DNS and the
  statistical model. A more quantitative agreement could, in
principle, be obtained by instead choosing $\alpha$ and $\kappa$ to
match relevant properties of flow correlation functions.

\begin{figure}[h!]
\hspace{0.2cm}
\begin{overpic}[width=5.1cm]{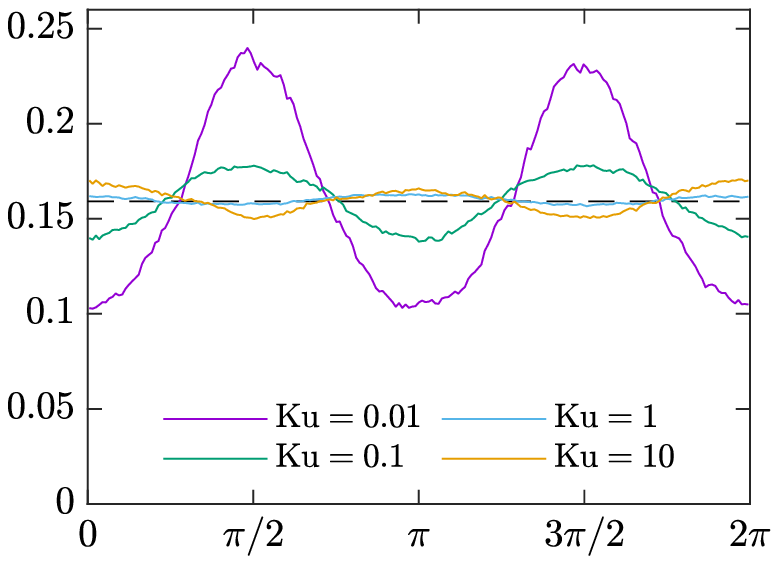}
\put(13,63){(a)}
\put(50,-6){$z/L$}
\put(-9,31){\rotatebox{90}{$P(z)$}}
\end{overpic}
\hspace{0.5cm}
\begin{overpic}[width=5.4cm]{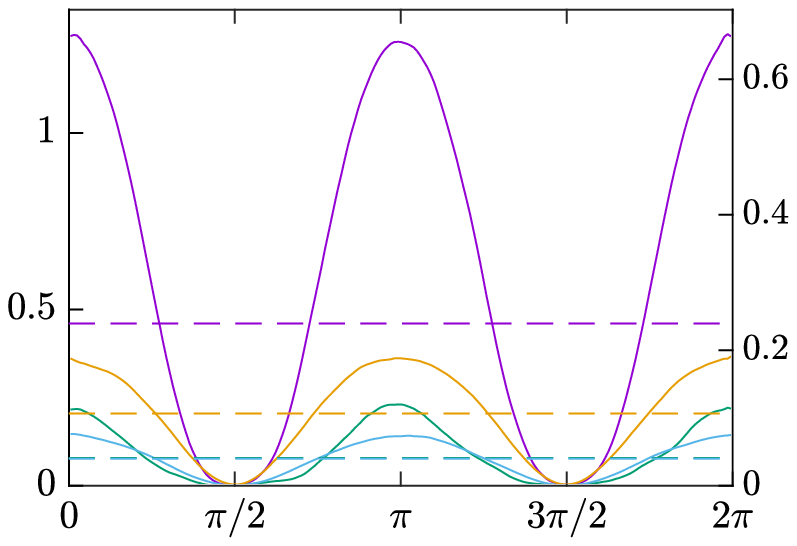}
\put(13,59){(b)}
\put(50,-6){$z/L$}
\put(-9,30){\rotatebox{90}{$\langle\ve U\cdot\ve n|z\rangle/u^\prime_{\rm rms}$}}
\put(101,60){\rotatebox{-90}{$\langle\ve U\cdot\ve n|z\rangle/U$}}
\end{overpic}
\hspace{0.9cm}
\begin{overpic}[width=5cm]{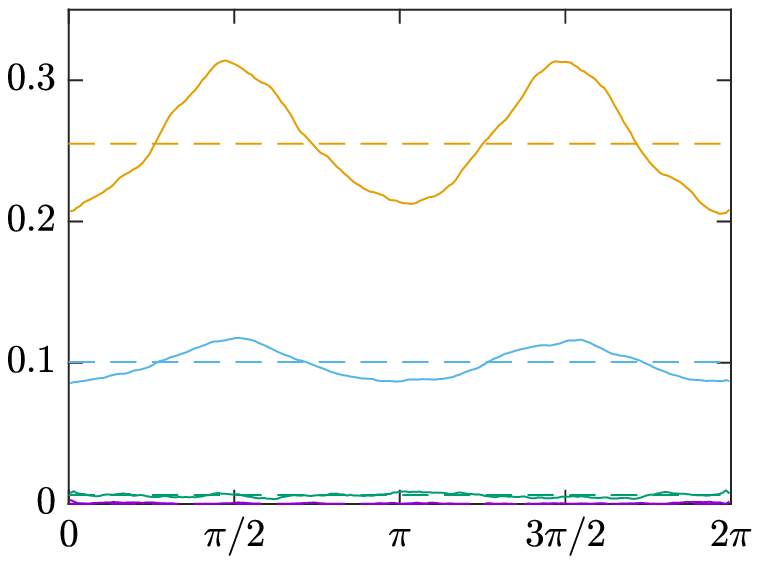}
\put(13,65){(c)}
\put(50,-6){$z/L$}
\put(-9,30){\rotatebox{90}{$\langle\ve u'\cdot\ve n|z\rangle/u^\prime_{\rm rms}$}}
\end{overpic}
\caption{\label{fig:4} (Color online) Microswimmer accumulation
  and alignment for three-dimensional statistical model simulations
  for different $\ku$. (a) distribution of $z$, (b) average $\ve
  U\cdot\ve n$ conditional on $z$, and (c) average $\ve u'\cdot\ve n$
  conditional on $z$.  Dashed horizontal lines show the averages
  (b) $\langle\ve U\cdot\ve n\rangle/u^\prime_{\rm rms}$ and
  (c) $\langle\ve u'\cdot\ve n\rangle/u^\prime_{\rm rms}$ without
  condition on $z$.  Parameters: $\Lambda=1$, $\alpha=1.92$,
  $\kappa=0.08$, and $\Phi=0.54$.}
\end{figure}

Figure~\ref{fig:4} shows spatial accumulation, $P(z)$, and
projection on the particle orientation on both the mean flow,
$\langle\ve U\cdot\ve n|z\rangle$, and the flow fluctuations,
$\langle\ve u'\cdot\ve n|z\rangle$, as functions of $z$.  We let
$\Phi=0.54$ and use $\alpha$ and $\kappa$ from our DNS.  For
$\ku=0.01$, Fig.~\ref{fig:4}(a) shows that particles accumulate
in high-shear regions, just as for the laminar Kolmogorov flow. The degree of accumulation is of the same
order as in Fig.~\ref{fig:1}(a) with small but non-zero inverse
P\'eclet numbers, $\Pe^{-1}$. Small deviations are
expected because the noise has different statistics and because the
swimming velocity is $v_\mathrm{s}/U=0.28$ in Fig.~\ref{fig:4}(a),
while it is $v_\mathrm{s}/U=0.2$ in Fig.~\ref{fig:1}(a).  Comparison of
Fig.~\ref{fig:4}(b) (right vertical scale) and
Fig.~\ref{fig:1}(b) shows the same degree of high alignment between
the mean flow velocity and the particle orientation, and
Fig.~\ref{fig:4}(c) shows that the alignment with the
stochastic part of the flow is negligible for small values of $\ku$.

For larger values of $\ku$, $\ku=10$, we compare to our DNS results.
Similar to the DNS in Fig.~\ref{fig:2}(c), the stochastic
fluctuations at large $\ku$ ruin the accumulation observed for small values of $\ku$ and particles are essentially uniformly
distributed.  Moreover, we find that $\langle\ve U\cdot\ve n|z\rangle$
and $\langle\ve u'\cdot\ve n|z\rangle$ in Fig.~\ref{fig:4}(b,c)
are both in qualitative agreement with the DNS data with $\Phi=0.54$
in Fig.~\ref{fig:3}(b,c), though the details do not match. The
alignment $\langle\ve U\cdot\ve n|z\rangle$ is approximately a factor
2 times larger in the statistical model, and $\langle\ve u'\cdot\ve
n|z\rangle$ is somewhat smaller and stronger modulated than the DNS
results.

In conclusion, our results for the statistical model qualitatively interpolates between the laminar Kolmogorov flow results for small values of $\ku$ and the DNS results for large values of $\ku$.
For intermediate values of $\ku$, the alignment $\langle\ve U\cdot\ve n|z\rangle$ has a minimum somewhere around $\ku\sim 1$, while $\langle\ve u'\cdot\ve n|z\rangle$ increases monotonously with $\ku$.

The results in Fig.~\ref{fig:4}(b,c) show that the alignment
conditional on $z$ is essentially obtained by a periodic modulation of
the unconditional alignment. As observed in the DNS
  simulations, alignment with the fluctuations is maximal around the
  zeroes of the mean flow, where the mean gradients are lager.  To
investigate how the unconditional alignment changes with the
parameters of the dynamics, we decompose the projection of the flow
velocity on the direction of the swimmer into the contributions from
the mean and the fluctuations (\ref{eq:decomp2}). As already
discussed, swimming breaks the fore-aft symmetry, allowing the average
alignment to take non-zero values.  For the case without the mean
flow, $U=0$, the perturbation theory of
Ref.~\cite{borgnino2019alignment} gives an estimate of $\langle\ve u^\prime\cdot\ve n\rangle$
and explains the alignment between $\ve u'$ and $\ve n$.  When a mean
flow is present, perturbation theory is much harder.  As we remarked in
Section~\ref{sec3}, the limit $\Phi\to 0$ is singular in the
deterministic dynamics.  Moreover, Fig.~\ref{fig:1} shows that as
$\Pe^{-1}$ approaches zero, the alignment abruptly
changes behavior, making the limit $\ku\to 0$ singular as well.
Therefore, instead of making a perturbation theory around the
deterministic solutions found in Section~\ref{sec3}, we consider a
perturbation theory in terms of small $\alpha=U/u^\prime_{\rm rms}$
and $\Phi=v_\mathrm{s}/u^\prime_{\rm rms}$ around the limit of tracer
particles advected by the stochastic part of the fluid velocity. In
this limit, we expect the second term in Eq.~(\ref{eq:decomp2}), $\langle\ve u^\prime\cdot\ve n\rangle$, to be
mainly unaffected compared to the $\alpha=0$ case studied in
Ref.~\cite{borgnino2019alignment}.  We argue that for small $\kappa$,
$\alpha$ and $\Phi$, the first term of Eq.~(\ref{eq:decomp2}) takes
the form
\begin{align}
\frac{\langle\ve U\cdot\ve n\rangle}{u^\prime_{\rm rms}}\sim {\mathcal C}_{\Lambda}\kappa^2\alpha^2\Phi\,,
\label{eq:projection_U_general}
\end{align}
where the coefficient ${\mathcal C}_\Lambda$ depends on the flow and the particle shape. The rationale for the above expression is as follows.

First, using the symmetry of the dynamics (\ref{eq:3}) and
(\ref{eq:4}) under the joint transformation $v_\mathrm{s}\to-v_\mathrm{s}$
and $\ve n\to-\ve n$, $\langle\ve u'\cdot\ve n\rangle$ must be an odd
function of $\Phi$, scaling linearly with $\Phi$ for small values of
$\Phi$~\cite{borgnino2019alignment}.  The same argument applies to
$\langle\ve U\cdot\ve n\rangle$, giving the linear scaling in
Eq.~(\ref{eq:projection_U_general}).  For the Kolmogorov mean flow,
$U_x=U\cos(z/L)$, the average projection on $\ve n$ is invariant under
reversing the sign of either the velocity field, $U\to-U$, or the box
length, $L\to-L$.  Hence, $\langle\ve U\cdot\ve n\rangle$ must be an
even function in both $U$ and $L$, implying it is also even in
$\alpha$ and $\kappa$.  When $\alpha=0$, there is no mean flow to
align with and the lowest contributing order in
Eq.~(\ref{eq:projection_U_general}) is thus quadratic.  When
$\kappa=0$, the mean flow is constant and, since a constant mean flow
is not physically distinguishable from a flow without a mean, the
alignment must vanish for this case, meaning that
Eq.~(\ref{eq:projection_U_general}) is quadratic to lowest order in
$\kappa$ in a physical flow. As noted in Sec.~\ref{sec:SM}, $\kappa\propto {\rm Re}^{-1/2}$, so
Eq.~(\ref{eq:projection_U_general}) implies that alignment with the
mean flow decreases with ${\rm Re}$, as indeed observed in our DNS
(see inset of Fig.~\ref{fig:3}(a)). However, the alignment decreases somewhat slower than predicted by the $\kappa^2$ scaling, indicating that the coefficient $C_\Lambda$ increases with $\Re$.
It is expected that $C_\Lambda$ depends on the statistics of the stochastic part of the flow, and hence on $\Re$ for the turbulence simulations and on $\ku$ in the statistical model.
Finally, the coefficient $C_\Lambda$ also depends on the shape factor $\Lambda$.
When $\Lambda=0$, the
swimmer becomes spherical and does not couple to the strain part of
the mean flow. We therefore expect that the coefficient ${\mathcal C}_\Lambda$ goes to zero as $\Lambda\to 0$.

\begin{figure}[h!]
\begin{overpic}[width=5.8cm]{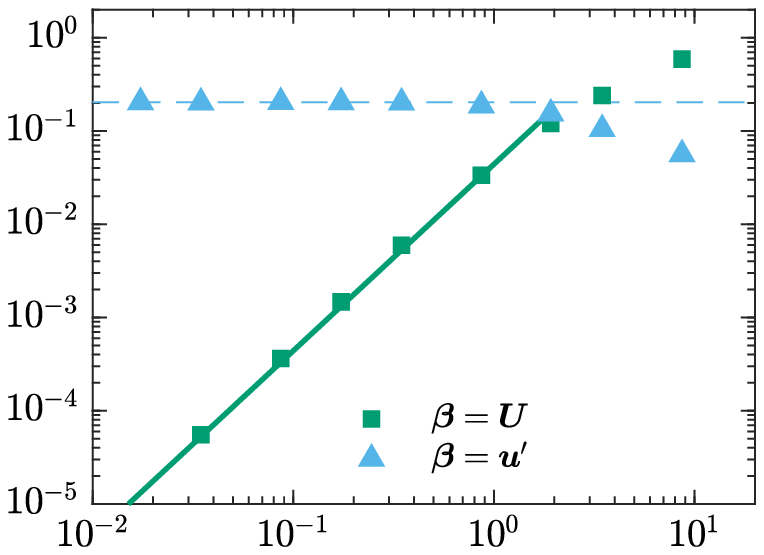}
\put(14,65){(a)}
\put(50,30){$\sim\alpha^2$}
\put(50,-6){$\alpha$}
\put(-8,32){\rotatebox{90}{$\langle\ve\beta\cdot\ve n\rangle/u^\prime_{\rm rms}$}}
\end{overpic}
\begin{overpic}[width=5.8cm]{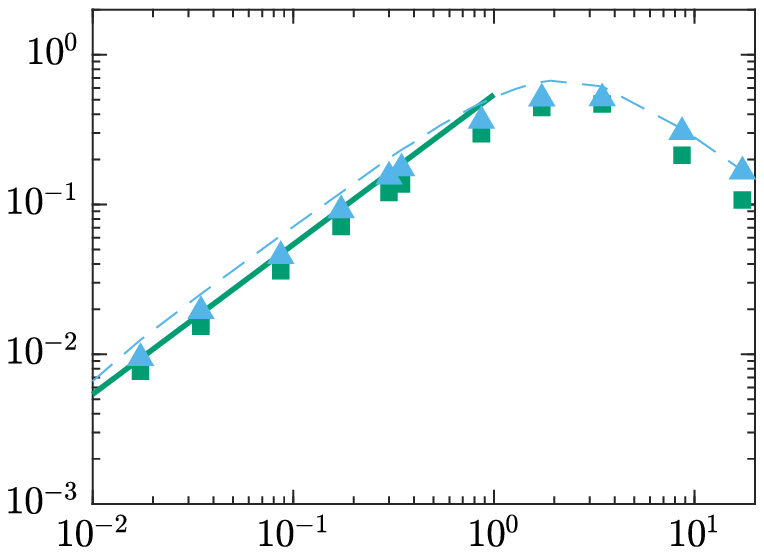}
\put(14,65){(b)}
\put(50,-6){$\Phi$}
\put(40,35){$\sim\Phi$}
\end{overpic}
\begin{overpic}[width=5.95cm]{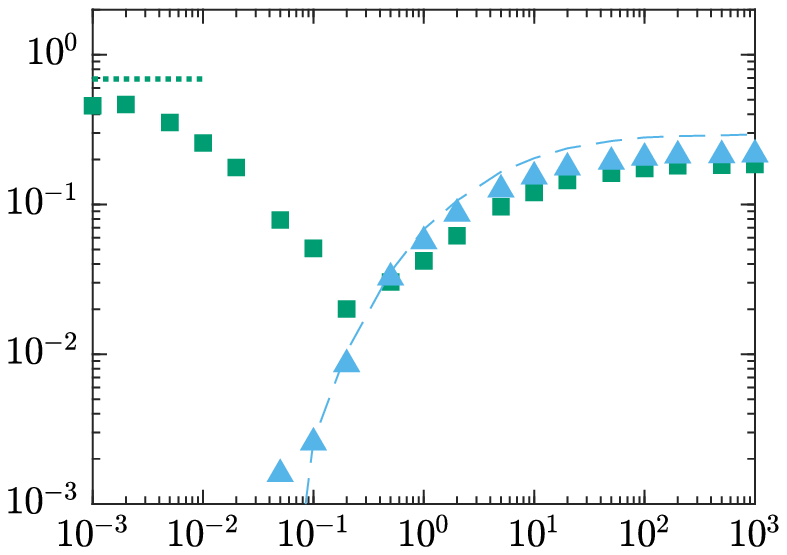}
\put(88,64){(c)}
\put(50,-6){$\ku$}
\end{overpic}
\caption{\label{fig:5} (Color online) Statistical model simulations of
  average projection $\langle\ve\beta\cdot\ve n\rangle$, with
  $\ve\beta=\ve U$ (green,$\Box$) and $\ve\beta=\ve u^\prime$
  (blue,$\vartriangle$).  The data is plotted against the dimensionless
  scale of (a) the mean flow velocity $\alpha$, (b) swimming speed $\Phi$
  and (c) the Kubo number $\ku$.  The solid green lines show
  Eq.~(\ref{eq:projection_U_general}) with a coefficient ${\cal
    C}_\Lambda=23$ fitted for small $\alpha$, $\Phi$ and $\kappa$ in
  panel (a).  Dashed blue lines show simulation results of $\langle
  u^\prime\cdot\ve n\rangle$ for the case of no mean flow, $\alpha=0$.
  The dotted green line in panel (c) shows results of $\langle\ve
  U\cdot\ve n\rangle$ in the deterministic limit
  [Eqs.~(\ref{eq:x1dot}--\ref{eq:nz1dot})] (converted to the
  dimensional units used here).  Parameters: $\Lambda=1$,
  $\kappa=0.08$, (a,b) $\ku=10$, (a,c) $\Phi=0.3$, and (b,c) $\alpha=1.92$.}
\end{figure}

Unlike DNS of the Kolmogorov flow, the statistical
  model allows to freely choose the dimensionless parameters $\alpha$
  and $\kappa$.  To verify the form (\ref{eq:projection_U_general}), we
therefore use statistical model simulations of rod-like swimmers,
$\Lambda=1$, in three spatial dimensions.
Fig.~\ref{fig:5}(a) shows results of $\langle\ve U\cdot\ve
n\rangle$ and $\langle\ve u'\cdot\ve n\rangle$ against $\alpha$ with
small $\Phi$ and $\kappa$.  The solid green line shows
Eq.~(\ref{eq:projection_U_general}) with a fitted prefactor ${\cal
  C}_\Lambda=23$ for our parameters $\ku=10$ and $\Lambda=1$.  We
observe that the predicted quadratic scaling of $\langle\ve U\cdot\ve
n\rangle$ with small $\alpha$ qualitatively agrees  with the simulation
data all the way up to the value $\alpha=1.92$ of the DNS. The
alignment with the stochastic part, $\langle\ve u'\cdot\ve n\rangle$,
is only weakly affected by the mean flow. For $\alpha<2$, it is
approximately equal to the value obtained without mean flow (dashed
blue line).

In Fig.~\ref{fig:5}(b) we show the same quantities against $\Phi$ for the value $\alpha=1.92$ of our DNS with $\Re=730$.
Even though $\alpha$ is not small, the symmetry argument after Eq.~(\ref{eq:projection_U_general}) anyway implies a linear scaling of $\langle\ve U\cdot\ve n\rangle$ with $\Phi$ for small values of $\Phi$, but the prefactor is somewhat smaller compared to Eq.~(\ref{eq:projection_U_general}) with ${\cal C}_\Lambda=23$ (solid green line)
The alignment $\langle\ve u'\cdot\ve n\rangle$ is approximately the same as for the case $\alpha=0$ (dashed blue line), showing a linear scaling for small $\Phi$ as predicted in Ref.~\cite{borgnino2019alignment}.
Similar to for Fig.~\ref{fig:4}, our results agree qualitatively with the results in the DNS if $\Phi<1$, but $\langle\ve U\cdot\ve n\rangle$ comes out larger and $\langle\ve u'\cdot\ve n\rangle$ comes out smaller than the DNS in Fig.~\ref{fig:3}.
For $\Phi$ larger than $\sim 1$, the alignment decreases in the statistical
model, while it increases in the DNS.
The reason for this difference is the presence of the inertial range
in a turbulent flow, with fluctuations over different scales and
times. Such range is not present in the statistical model where the
fluctuations take place on a single length scale.

Figure~\ref{fig:5}(c) shows how the alignment changes from
the limit of laminar Kolmogorov flow at small $\ku$ to the limit of a
turbulent Kolmogorov flow at large $\ku$.  Since $\tau_{\rm f}$ is not
a relevant time scale in the latter limit, plateaus are formed in both
alignments $\langle\ve U\cdot\ve n\rangle$ and $\langle\ve
u'\cdot\ve n\rangle$ for large $\ku$.  When $\ku=0$, the flow is
deterministic and the dynamics discussed in Section~\ref{sec3}
applies. The green dotted line in Fig.~\ref{fig:5}(c) shows
simulation results of $\langle\ve U\cdot\ve n\rangle$ in the
deterministic limit [Eqs.~(\ref{eq:x1dot}--\ref{eq:nz1dot})], scaled
to the units used in this section. The numerical data indicates that
the deterministic limit is singular: the average with $\ku=0$ is not
approached smoothly for small but finite values of $\ku$. This is
consistent with the behavior in Fig.~\ref{fig:1} (b), where a small
noise abruptly changes the spatial dependence of the alignment.  As
observed in Fig.~\ref{fig:4}(b), $\langle\ve U\cdot\ve
n\rangle$ is non-monotonic, having a large alignment
for both small and large values of $\ku$ with a minimum around
$\ku\sim 1$.  The mechanism is however different in the two limits,
for small $\ku$ the alignment is purely determined by the mean flow
$\ve U$, while in the limit of large $\ku$ the contributions from the
mean flow and the stochastic part are of the same order.  The
alignment $\langle\ve u'\cdot\ve n\rangle$ is approximately given by
the case without mean flow (dashed blue line).

In summary, the numerical data of the statistical model show a
linear scaling with $\Phi$ and a quadratic scaling with $\alpha$, as predicted
by Eq.~(\ref{eq:projection_U_general}).
The $\kappa^2$ scaling is hard to verify with $\alpha=1.92$ using the
statistical model. We instead verify this scaling for the case of smaller
values of $\alpha$ below.

\subsection{Perturbation theory in two spatial dimensions}
To understand the physical mechanisms that give rise to the alignment,
we evaluate ${\mathcal C}_\Lambda$ analytically in the statistical
model for small values of $\ku$ in two spatial dimensions.  To this
end we make a series expansion around the limit where the dynamics is
dominated by the stochastic part of the velocity field (i.e. small $\alpha$).  In addition we consider both
  $\kappa$ and $\Phi$ small consistently with the limit in
  Eq.~(\ref{eq:projection_U_general}), and we assume that the Kubo
  number is small in order to close the expansion at second order in
  $\ku$.  The computation is done with the further assumption that
both $\alpha$ and $\Phi$ are small enough compared to $\ku$, so that
the periodic trajectories in the deterministic dynamics can be
neglected.  The perturbation theory follows the method introduced in
Refs.~\cite{Gustavsson2011Ergodic,Gustavsson2014Clustering,gustavsson2016gyrotaxis,Gustavsson2017Statistical}
and reviewed in~\cite{Gustavsson2016}, but with one important
difference: we expand the angular dynamics around the solution
$\varphi^{(0)}(t)$  obtained if the mean flow
gradients are set to zero and if the stochastic part of the flow
gradients are evaluated at a fixed position.  Even though
$\varphi^{(0)}(t)$ is a function of $\ku$, we do not expand
trigonometric functions of $\varphi^{(0)}(t)$ in terms of small
$\ku$. As a result, we avoid secular terms and the resulting solution
converges for large times.  The details of the expansion are given in
Appendix~\ref{app:perturbation}.  The general result from our
expansion is obtained in Eq.~(\ref{eq:appendix_Un_General_result}) and
reported below
\begin{align}
\label{eq:projection_U}
\frac{\langle\ve U\cdot\ve n\rangle}{u^\prime_{\rm rms}}&=
\frac{1}{2}\Phi\Lambda\int\frac{{\rm d}^2\ve y}{(2\pi L)^2}{\rm Tr}(\ma S(\ve y)^2)\int_0^t{\rm d}t_1\int_0^{t_1}{\rm d}t_2
\\&\times
\exp\left[
-\frac{2+\Lambda^2}{8}\left[
\int_{t_2}^t{\rm d}t'\int_{t_2}^t{\rm d}t''C(t'-t'')
+\int_{t_2}^{t_1}{\rm d}t'\int_{t_2}^{t_1}{\rm d}t''C(t'-t'')
+2\int_{t_2}^t{\rm d}t'\int_{t_2}^{t_1}{\rm d}t''C(t'-t'')
\right]
\right]\,.\nonumber
\end{align}
This expression applies to a general mean flow that is either periodic
or zero at the boundaries of a box of side length $2\pi L$. It
consists of one spatial average over the squared mean-flow strain,
${\rm Tr}(\ma S(\ve y)^2)$, where $\ma S$ has components
$S_{ij}=(\partial_jU_i+\partial_iU_j)/2$, multiplied by time integrals
of the correlation function of the stochastic fluid gradients
evaluated at a fixed position $\ve x_0$, $C(t)=\langle
\partial_ju'_i(\ve x_0,t)\partial_ju'_i(\ve x_0,0)\rangle$.  In the
derivation of Eq.~(\ref{eq:projection_U}), the mean part of the flow
has been evaluated at fixed positions (instead of Lagrangian
trajectories), which is a valid approximation for small values of
$\ku$, but may fail when $\ku$ is large.

Equation~(\ref{eq:projection_U}) converges in the limit of large time.
In contrast, perturbation theory of angular dynamics in terms of small $\ku$ in previous studies gave rise to secular terms~\cite{Gustavsson2014Clustering,gustavsson2016gyrotaxis,Gustavsson2017Statistical}.
In these cases a self-consistency condition was used to remove the secular terms, leading to infinite recursion relations that can only be solved in limiting cases.
We expect that the solution method introduced here will apply to a large number of problems that were not accessible using the old method, including solving the angular dynamics in Refs.~\cite{Gustavsson2014Clustering,gustavsson2016gyrotaxis,Gustavsson2017Statistical} without reference to infinite recursion relations.

Using the correlation function of the stochastic model allows to
evaluate Eq.~(\ref{eq:projection_U}) to lowest contributing order in
$\ku$ (see Appendix~\ref{app:perturbation} for details)
\begin{align}
\frac{\langle\ve U\cdot\ve n\rangle}{u^\prime_{\rm rms}}=\frac{\Lambda}{2(2+\Lambda^2)^2\ku^2}\kappa^2\alpha^2\Phi\,.
\label{eq:projection_U_smallKu}
\end{align}
This expression is on the form (\ref{eq:projection_U_general}) with
${\mathcal C}_\Lambda=\Lambda/(2(2+\Lambda^2)^2\ku^2)$.  We remark
that since $\alpha$ and $\Phi$ are assumed to be small enough compared
to $\ku$, the inverse dependence on $\ku^2$ in
Eq.~(\ref{eq:projection_U_smallKu}) does not pose a problem to the expansion.

It is instructive to analyze what mechanisms contribute to
Eq.~(\ref{eq:projection_U}).  First, we find that the flow velocity in
the translational dynamics~(\ref{eq:3}) does not contribute, and to
order $\ku^2$ the translational dynamics is implicitly determined by
the orientation through $\ve x(t) =\ve x_0+\int_0^t{\rm d}t_1
v_\mathrm{s} \ve n(t_1)$.  As a consequence, displacement due to the
mean flow does not contribute to Eq.~(\ref{eq:projection_U}).  The
displacement due to swimming affects the sampling of the factor $\ve
U$ in $\langle\ve U\cdot\ve n\rangle$, but also the factor $\ve n$ due
to a feed-back from swimming on the sampling of mean-flow gradients in
the rotational dynamics~(\ref{eq:4}).  If the displacement due to
swimming were not taken into account, the rotational dynamics would
essentially originate from Jeffery's orbits driven by time-dependent fluid
gradients.  But in that case $\langle\ve U\cdot\ve n\rangle$ must
vanish.  The feedback due to displacement from swimming shows that the
alignment with the mean flow is a more intricate effect.

Second, the vorticity of the mean flow does not contribute to
Eq.~(\ref{eq:projection_U}).  The alignment is instead determined by
the mean-flow strain, and there is a net alignment in any flow where
the spatial average in Eq.~(\ref{eq:projection_U}) is non-zero.  On
the other hand, in the contribution from the stochastic part of the
flow to Eq.~(\ref{eq:projection_U}), vorticity is the dominant
contribution for all values of $\Lambda$, and the sole contribution
for small values of $\Lambda$.  This is simply a consequence of the
strain being multiplied by the shape factor $\Lambda$ in Jeffery's
orientational dynamics~(\ref{eq:4}) and by a cancellation between
strain components multiplying different trigonometric functions of the
angle.  We conclude by observing that in Fig.~\ref{fig:4}(b),
swimmers show the largest alignment where the mean flow is high and
the flow gradients are small.  It may therefore be perceived as
surprising that it is the mean strain gradients rather than the mean
flow that determines the mean alignment.

Third, we find that preferential sampling of the stochastic part of
the flow does not contribute to Eq.~(\ref{eq:projection_U}), implying
that the stochastic flow and its gradients can be evaluated at a fixed
position.  This is in contrast to the alignment with the stochastic
part of the flow.  We have made an expansion similar to that leading
to Eq.~(\ref{eq:projection_U}), but for the stochastic part,
$\langle\ve u'\cdot\ve n\rangle$. This expansion shows that
$\langle\ve u'\cdot\ve n\rangle$ is not affected by the mean flow to
lowest contributing order in $\ku$.  Without a mean flow, the
alignment was derived for small $\ku$ in
Ref.~\cite{borgnino2019alignment}:
\begin{align}
\frac{\langle\ve u'\cdot\ve n\rangle}{u^\prime_{\rm rms}}=\Phi\Lambda\ku^2\,.
\label{eq:projection_uprim_smallKu}
\end{align}
As discussed in Ref.~\cite{borgnino2019alignment}, this alignment requires preferential sampling of the stochastic fluctuations of the flow.

Although Eq.~(\ref{eq:projection_U}) is derived for small values of
$\ku$, we use the scaling $\langle\ve U\cdot\ve
n\rangle\sim\alpha(\kappa\alpha)(\kappa\Phi)$ in
Eq.~(\ref{eq:projection_U_general}) to argue that the same mechanisms
are, at least for small $\alpha$ and $\Phi$, dominant in general flows
dominated by fluctuations, such as our DNS.  The first factor $\alpha$
trivially comes from $\ve U$ in $\langle\ve U\cdot\ve n\rangle/u_{\rm
  rms}$.  There are two ways $\kappa$ can enter the dynamics: either
from derivatives of the mean flow, which are proportional to
$\kappa\alpha$, or from the feedback from the swimming on the location
of the swimmer in the mean flow, which gives a factor $\kappa\Phi$.
We therefore expect the flow in the translational dynamics to be
negligible compared to the swimming, and the swimming to have a
feedback on the rotational dynamics, just as for small $\ku$ above.

\begin{figure}[h!]
\centering
\begin{overpic}[width=0.3\textwidth]{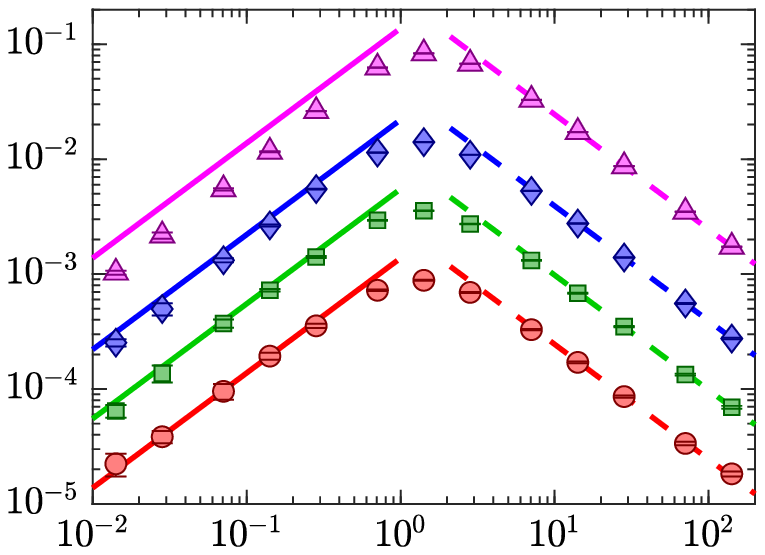}
\put(14,64){(a)}
\put(50,-6){$\Phi$}
\put(-8,30){\rotatebox{90}{$\langle\ve U\cdot\ve n\rangle/u^\prime_{\rm rms}$}}
\end{overpic}
\hspace{0.5cm}
\begin{overpic}[width=0.3\textwidth]{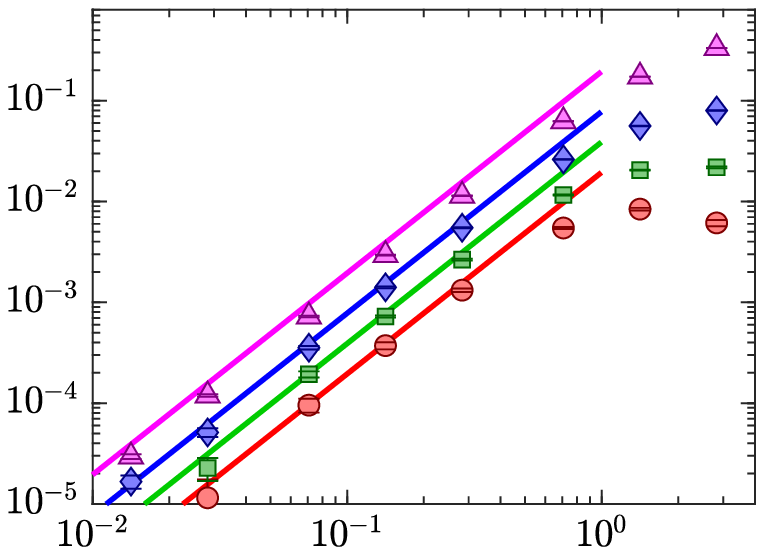}
\put(14,64){(b)}
\put(50,-6){$\alpha$}
\put(-8,30){\rotatebox{90}{$\langle\ve U\cdot\ve n\rangle/u^\prime_{\rm rms}$}}
\end{overpic}
\hspace{0.5cm}
\begin{overpic}[width=0.3\textwidth]{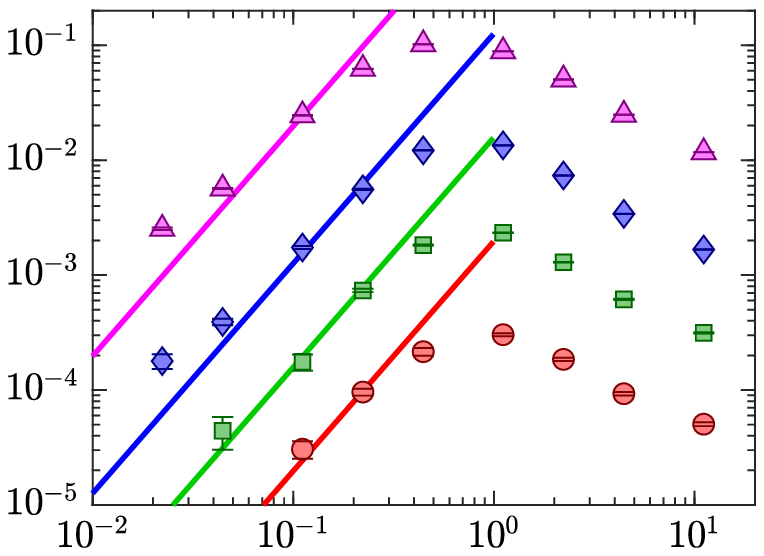}
\put(14,64){(c)}
\put(50,-6){$\kappa$}
\put(-8,30){\rotatebox{90}{$\langle\ve U\cdot\ve n\rangle/u^\prime_{\rm rms}$}}
\end{overpic}
\\\vspace{0.5cm}
\begin{overpic}[width=0.3\textwidth]{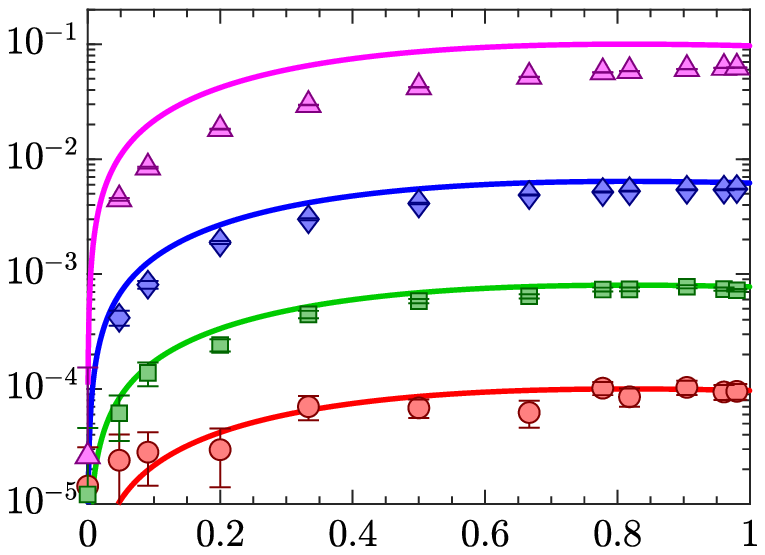}
\put(14,64){(d)}
\put(50,-6){$\Lambda$}
\put(-8,30){\rotatebox{90}{$\langle\ve U\cdot\ve n\rangle/u^\prime_{\rm rms}$}}
\end{overpic}
\hspace{0.5cm}
\begin{overpic}[width=0.307\textwidth]{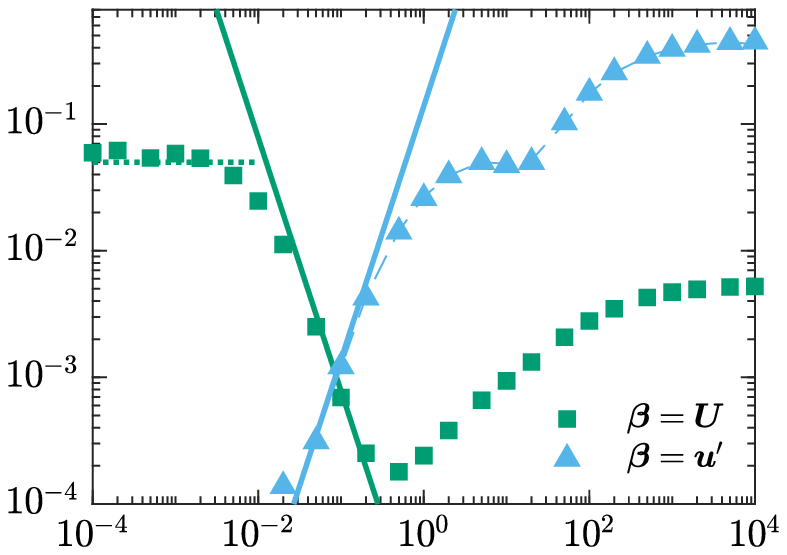}
\put(14,62){(e)}
\put(50,-6){$\ku$}
\put(-8,30){\rotatebox{90}{$\langle\ve\beta\cdot\ve n\rangle/u^\prime_{\rm rms}$}}
\end{overpic}
\caption{\label{fig:6}
(Color online) Test of analytical predictions.
  (a--d) Average projection of $\ve U$ onto the direction $\ve n$ plotted against (a) the swimming speed $\Phi$, (b) mean flow magnitude $\alpha$, (c) wave number $\kappa$, and (d) shape factor $\Lambda$ for small $\ku$.
Numerical data from the stochastic model in two spatial dimensions are plotted as markers.
Solid lines show the theory (\ref{eq:projection_U_smallKu}) and dotted lines in panel (a) show the large-$\Phi$ theory~(\ref{eq:projection_U_LargePhi}).
Panel (e) shows data corresponding to Fig.~\ref{fig:5}(c), but in two spatial dimensions with small $\alpha$ and $\Phi$.
The solid lines show the theory (\ref{eq:projection_U_smallKu}) [green] and (\ref{eq:projection_uprim_smallKu}) [blue].
Panel (a): $\alpha=0.071$ (red,$\circ$), $\alpha=0.14$ (green,$\Box$), $\alpha=0.28$ (blue,$\Diamond$), and $\alpha=0.71$ (magenta,$\vartriangle$).
Panel (b): $\Phi=0.071$ (red,$\circ$), $\Phi=0.14$ (green,$\Box$), $\Phi=0.28$ (blue,$\Diamond$), and $\Phi=0.71$ (magenta,$\vartriangle$).
Panels (c,d): $\alpha=\Phi=0.071$ (red,$\circ$), $\alpha=\Phi=0.14$ (green,$\Box$), $\alpha=\Phi=0.28$ (blue,$\Diamond$), and $\alpha=\Phi=0.71$ (magenta,$\vartriangle$).
Panel (e): $\alpha=\Phi=0.14$.
Panels (a--c,e): $\Lambda=0.98$.
Panels (a--b,d--e): $\kappa=0.22$.
Panels (a--d): $\ku=0.1$.
}
\end{figure}

\subsection{Comparison of analytical theory with numerical simulations}
Equation~(\ref{eq:projection_U_smallKu}) is compared to numerical
simulations with small values of $\ku$ in
Fig.~\ref{fig:6}.  We observe good agreement in
general, but there is a systematic trend of the theory to overpredict
the alignment. A plausible explanation for this is that these
deviations come from a higher-order additive contribution.

Figures~\ref{fig:6}(a--c) show that the predicted
scaling $\langle\ve U\cdot\ve n\rangle/u^\prime_{\rm rms}\sim\kappa^2\alpha^2\Phi$ agrees
well with the theory~(\ref{eq:projection_U_smallKu}) for small enough
$\Phi$, $\alpha$, and $\kappa$.  For large $\Phi$, panel (a) also shows
the theory
\begin{align}
\frac{\langle\ve U\cdot\ve n\rangle}{u^\prime_{\rm rms}}=\frac{\Lambda\alpha^2}{2\Phi}
\label{eq:projection_U_LargePhi}
\end{align}
for large swimming speeds (dashed lines), which is derived in
Appendix~\ref{sec:large_Phi}. This expression is valid if $\Phi$ is
large compared to $\alpha$ and unity. Moreover, it does not depend on
the Kubo number and thus it applies also to
Fig.~\ref{fig:5}(b), up to a change in the prefactor due to the spatial dimension (not shown).  The theory
(\ref{eq:projection_U_LargePhi}) works well for the statistical model,
but it is expected to fail in the DNS due to long-range spatial
correlations in the inertial range, not present in the statistical
model, c.f. Fig~\ref{fig:3}. Figure~\ref{fig:6}(d)
shows the shape dependence of the alignment for small $\ku$. The
theory agrees well if $\alpha$ and $\Phi$ are small.
Equation~(\ref{eq:projection_U_smallKu}) shows that $\langle\ve
U\cdot\ve n\rangle$ is linear in $\Lambda$ for small values of
$\Lambda$, and approximately independent of $\Lambda$ for
$\Lambda\gtrsim 0.5$ where it reaches $90\%$ of its maximal
value. This further motivates our choice of $\Lambda=1$ in the DNS
simulations, at least for small values of $\ku$.  Finally,
Fig.~\ref{fig:6}(e) shows how the alignment depends on
$\ku$.  The theory agrees for small values of $\ku$, but fails for the
deterministic part when $\ku$ becomes very small, $\ku\sim0.01$.  The
reason is that periodic orbits are formed when $\ku$ is not large
enough compared to $\Phi$ and $\alpha$, which are not included in the
theory.  For very large values of $\ku$, the system obtains very large
alignment.  This is a result of periodic orbits of elongated swimmers
in frozen two-dimensional stochastic flows.  This effect is not
present in the 3D simulations in Fig.~\ref{fig:5}(c)
because the dynamics of elongated swimmers in 3D frozen flows is
chaotic.  Equations~(\ref{eq:projection_U_smallKu}) and
(\ref{eq:projection_uprim_smallKu}) are equal at the scale $\ku_{\rm
  c}=2^{-1/4}\sqrt{\kappa\alpha/(2+\Lambda^2)}$ for small values of
$\kappa$, $\alpha$ and $\ku$. Below this scale, alignment is dominated
by the mean part of the flow and, above it, the alignment with the
stochastic part of the flow dominates.  This scale also explains the
locations of the cross-over in the 3D simulations with larger
$\alpha$, $\ku_{\rm c}\approx 0.2$, consistent with the results in
Fig.~\ref{fig:5}(c).  We conclude by remarking that the
alignment with the stochastic part of the velocity in
Fig.~\ref{fig:6}(e) is indistinguishable to the case
with $\alpha=0$ (dashed blue line), consistent with our theory for the
stochastic part that is independent of $\alpha$ to lowest order in
$\ku$.


\section{Conclusions}
\label{sec:sec5}

We investigated the statistics of accumulation and alignment of
elongated microswimmers in an inhomogeneous Kolmogorov flow.
We used different techniques to address three specific situations.
In the case of a laminar, steady flow, we used a dynamical systems approach
together with numerical simulations to study the effect of a small
rotational diffusivity. In this case, we observe a robust alignment
of the swimming direction with the underlying flow together with a
non-homogeneous distribution of swimmers which tend to accumulate in
regions of high shear.

The case of turbulent Kolmogorov flow is studied by means of DNS of the Navier-Stokes equations. In this case,
we find that swimmers are still able to align with the direction of
the mean flow in spite of the presence of turbulent fluctuations.
Interestingly, we observe an alignment also with the local turbulent
velocity fluctuations, which confirm the results recently obtained
for an homogeneous isotropic flow \cite{borgnino2019alignment}.
On the contrary, the accumulation of swimmers in high shear regions
is disrupted by the presence of turbulent fluctuations for all the
values of the parameters investigated.

In order to interpolate between these two regimes, we also
considered a steady Kolmogorov flow with an overlying stochastic
component. In this case we are able to develop an analytic perturbative
approach which describes the origin of the observed alignment in a
variety of limits.
Remarkably, the alignment with the mean flow
finds its roots in the mean strain and the fluctuating vorticity. In contrast the alignment with the stochastic part of the flow is not affected by the mean flow but it is determined by the fluctuating strain.

Our results indicate that elongated microswimmers are able to align their
swimming direction with the mean flow in a wide range of conditions and we
expect that a similar behavior should be observable in other shear
flows. For example Eq.~(\ref{eq:projection_U_general}), together with the
definitions of $\kappa$ and $\alpha$, suggests that the alignment-induced contribution of the mean flow to the displacement of swimmers can be
estimated by $\langle \bm U\cdot \bm n\rangle\approx C\tau_\eta^2S^2 v_\mathrm{s}$, with $S$ the local shear rate and $C$ a
Reynolds-dependent parameter that is O(10) for our flow
(see section
\ref{sec5}). In the presence of strong stratification,
vertical structures of velocity profiles are observed in the ocean at a scale of about $5$--$10\,
{\rm m}$, with shear rates of order $S= 0.01\, {\rm s}^{-1}$, corresponding to typical
velocities in the range $5$--$10\, {\rm cm/s}$ \cite{shroyer2014stratification,
durham2012thin}. 
In the ocean, in weakly turbulent conditions, one can have $\epsilon\sim 10^{-9}\,\mathrm J\,\mathrm{kg}^{-1}\,\mathrm s^{-1}$ \cite{franks2022oceanic,thorpe2007introduction}, giving $\tau_\eta^2\sim 10^3 \mathrm s^2$. For such values of $\epsilon$ typical $u^\prime_{\rm rms}$ range between $0.3$--$1\, {\rm cm/s}$ \cite{yamazaki1996comparison}.
Based on these estimates one gets $\langle \bm U\cdot \bm n\rangle\sim O(v_\mathrm{s})$. Single-celled motile organisms can swim at speeds up to $v_\mathrm{s}=200$--$300 \mu{\rm m/s}$. Some phytoplankters can form elongated
chains of up to 10 cells which can reach $v_\mathrm{s}=1.5\, {\rm mm/s}$
\cite{lovecchio2019chain}.
 Thus, the alignment with the mean flow can impact the dispersion properties
of microswimmers in weak turbulence also in the absence of accumulation, widening the previous
observations made in steady flows \cite{dehkharghani2019bacterial}.

\begin{acknowledgments}
We are indebted to B. Mehlig for the numerous
  fruitful discussions.  MB, GB and FDL acknowledge support by the
  {\it Departments of Excellence} grant (MIUR). CINECA is
  acknowledged for computing resources, within the INFN-Cineca
  agreement INFN-fieldturb.
KG acknowledges support by Vetenskapsrådet, Grant No. 2018-03974.
  Statistical model simulations were performed on resources provided by the Swedish National Infrastructure for Computing (SNIC).
\end{acknowledgments}

\appendix

\section{Perturbation expansion in small Kubo numbers}
\label{app:perturbation}

To make a perturbation expansion in terms of small values of $\ku$, we dedimensionalize the dynamics (\ref{eq:1}-\ref{eq:4}) in terms of the flow scales in the stochastic model in Section \ref{sec:SM}, i.e. time is dedimensionalized by $\tau_{\rm f}$, space by $\ell_{\rm f}$ and velocity by $u_{\rm f}$.
In these units the dynamics in two spatial dimensions becomes
\begin{align}
\begin{split}
\dot{\ve x}_t&=\ku[\ve u(\ve x_t,t)+\Phi\ve n_t]\,,\\
\dot\varphi_t&=\ku[\ve n^\perp_t]^{\rm T}\ma B(\ve x_t,t)\ve n_t\,,
\end{split}
\label{eq:eomKu}
\end{align}
where $\ve n^\perp=(-\sin\varphi,\cos\varphi)$ is a unit vector perpendicular to $\ve n$ and $\ma B=\ma O+\Lambda\ma S$, with $\ma O=(\ma A-\ma A^{\rm T})/2$ and $\ma S=(\ma A+\ma A^{\rm T})/2$ being the antisymmetric and symmetric part of the fluid gradient matrix $\ma A$.
We use subscripts of $t$ to denote functional dependence on time, for example $\ve x_t=\ve x(t)$.
In this appendix we denote the mean and stochastic part of the flow and its gradients evaluated at the initial position $\ve x_0$ by
\begin{align}
\ve u(\ve x_0,t)=\ve U(\ve x_0)+\ve u(t)\hspace{0.5cm}\mbox{and}\hspace{0.5cm}\ma A(\ve x_0,t)=\ma A(\ve x_0)+\ma A(t)\,,
\end{align}
i.e. a spatial argument denotes the mean part of the flow and a time argument the stochastic part of the flow at position $\ve x_0$.
By decomposing $\ma B(\ve x_t,t)$ in terms of its stochastic part at the initial position $\ve x_0$, $\ma B(t)$, and the remainder $\ma B(\ve x_t,t)-\ma B(t)$, we write the implicit solution to Eq.~(\ref{eq:eomKu}) as
\begin{align}
\begin{split}
\ve x_t&=\ve x_0+\ku\int_0^t{\rm d}t_1\left[\ve u(\ve x_{t_1},t_1)+\Phi\ve n_{t_1}\right]\,,
\\
\varphi_t&=\varphi^{(0)}_t+\ku\int_0^t{\rm d}t_1\left[
-[\ve n^{\perp,(0)}_{t_1}]^{\rm T}\ma B(t_1)\ve n^{(0)}_{t_1}
+[\ve n^\perp_{t_1}]^{\rm T}\ma B(\ve x_{t_1},t_1)\ve n_{t_1}\right]\,.
\end{split}
\label{eq:SmallKuImplicitSolution}
\end{align}
Here $\varphi^{(0)}_t$ is the solution to
\begin{align}
\dot\varphi^{(0)}_t=\ku[\ve n^{\perp,(0)}_t]^{\rm T}\ma B(t)\ve n^{(0)}_t\,,
\label{eq:phi0Dot}
\end{align}
with $\ve n^{(0)}_t=(\cos\varphi^{(0)}_t,\sin\varphi^{(0)}_t)$, starting from an initial angle $\varphi^{(0)}_0=\varphi_0$.
We make the following ansatz for the solution to second order in $\ku$
\begin{align}
\begin{split}
\ve x_t&=\ve x_0+\ku\ve x^{(1)}_t+\ku^2\ve x^{(2)}_t\,,
\\
\varphi_t&=\varphi^{(0)}_t+\ku\varphi^{(1)}_t+\ku^2\varphi^{(2)}_t\,.
\end{split}
\label{eq:ansatzSmallKu}
\end{align}
By recursively substituting $\ve x_t$ and $\varphi_t$ in Eq.~(\ref{eq:SmallKuImplicitSolution}) into the right hand side and discarding terms of third order or higher in $\ku$, we identify the expansion coefficients in the ansatz
\begin{align}
\begin{split}
\ve x^{(1)}_t&=\int_0^t{\rm d}t_1[\ve u(\ve x_{0},t_1)+\Phi\ve n^{(0)}_{t_1}]
\\
\ve x^{(2)}_t&=\int_0^t{\rm d}t_1[\ma A(\ve x_0,t_1)\ve x^{(1)}_{t_1}+\Phi\varphi^{(1)}_{t_1}\ve n^{\perp,(0)}_{t_1}]
\\
\varphi^{(1)}_t&=
\int_0^t{\rm d}t_1
[\ve n^{\perp,(0)}_{t_1}]^{\rm T}\ma B(\ve x_0)\ve n^{(0)}_{t_1}
\\
\varphi^{(2)}_t&=
\int_0^t{\rm d}t_1\big(
[\ve n^{\perp,(0)}_{t_1}]^{\rm T}[(\ve x^{(1)}_{t_1}\cdot\ve\nabla)\ma B(\ve x_{0},t_1)]\ve n^{(0)}_{t_1}
+\varphi^{(1)}_{t_1}[\ve n^{\perp,(0)}_{t_1}]^{\rm T}\ma B(\ve x_{0},t_1)\ve n^{\perp,(0)}_{t_1}
-\varphi^{(1)}_{t_1}[\ve n^{(0)}_{t_1}]^{\rm T}\ma B(\ve x_{0},t_1)\ve n^{(1)}_{t_1}\big)
\end{split}
\label{eq:solutionSmallKu}
\end{align}

\subsubsection{Alignment with the mean flow}
We start by evaluating the projection of the mean flow on the particle orientation.
Using the solution~(\ref{eq:solutionSmallKu}) and expanding to second order in $\ku$ gives
\begin{align}
\begin{split}
\ve U(\ve x_t)\cdot\ve n_t
&=
\ve U(\ve x_0)\cdot\ve n^{(0)}_t
+\ku\left(\varphi^{(1)}_t\ve U(\ve x_0)\cdot\ve n^{\perp,(0)}_t+[\ve n^{(0)}_t]^{\rm T}\ma A(\ve x_0)\ve x^{(1)}_t\right)
+\ku^2\Big(\varphi^{(2)}_t\ve U(\ve x_0)\cdot\ve n^{\perp,(0)}_t
\\&\!\!
-\frac{1}{2}[\varphi^{(1)}_t]^2\ve U(\ve x_0)\cdot\ve n^{(0)}_t
+\varphi^{(1)}_t[\ve n^{\perp,(0)}_t]^{\rm T}\ma A(\ve x_0)\ve x^{(1)}_t
+[\ve n^{(0)}_t]^{\rm T}\ma A(\ve x_0)\ve x^{(2)}_t+\frac{1}{2}[\ve n^{(0)}_t]^{\rm T}[(\ve x^{(1)}_t\cdot\ve\nabla)\ma A(\ve x_0)]\ve x^{(1)}_t
\Big)\,.
\end{split}
\label{eq:Un_start}
\end{align}
As discussed in Section~\ref{sec4}, the average $\langle\ve U(\ve x_t)\cdot\ve n_t\rangle$ must be an odd function in $\Phi$ due to the symmetries of the dynamics~\cite{borgnino2019alignment}.
Keeping only odd terms in $\Phi$ after inserting $\ve x^{(i)}_t$ and $\ve\varphi^{(i)}_t$ into Eq.~(\ref{eq:Un_start}), the average projection of $\ve U$ on $\ve n$ becomes
\begin{align}
\begin{split}
\langle\ve U(\ve x_t)\cdot\ve n_t\rangle
&=\ku\Phi\int_0^t{\rm d}t_1\langle[\ve n^{(0)}_t]^{\rm T}\ma A(\ve x_0)\ve n^{(0)}_{t_1}\rangle
\\&
+\frac{1}{2}\ku^2\Phi\int_0^t{\rm d}t_1\int_0^{t_1}{\rm d}t_2\Big\langle
2[\ve n^{\perp,(0)}_{t_1}]^{\rm T}[(\ve n^{(0)}_{t_2}\cdot\ve\nabla)\ma B(\ve x_0,t_1)]\ve n^{(0)}_{t_1}(\ve U(\ve x_0)\cdot\ve n^{\perp,(0)}_t)
\\&
\hspace{2.5cm}
+[\ve n^{(0)}_t]^{\rm T}[(\ve n^{(0)}_{t_1}\cdot\ve\nabla)\ma A(\ve x_0)]\ve u(\ve x_{0},t_2)
+[\ve n^{(0)}_t]^{\rm T}[(\ve n^{(0)}_{t_2}\cdot\ve\nabla)\ma A(\ve x_0)]\ve u(\ve x_{0},t_1)
\\&
\hspace{2.5cm}
+[\ve n^{(0)}_t]^{\rm T}[(\ve u(\ve x_{0},t_1)\cdot\ve\nabla)\ma A(\ve x_0)]\ve n^{(0)}_{t_2}
+[\ve n^{(0)}_t]^{\rm T}[(\ve u(\ve x_{0},t_2)\cdot\ve\nabla)\ma A(\ve x_0)]\ve n^{(0)}_{t_1}
\\&
\hspace{2.5cm}
+2[\ve n^{\perp,(0)}_{t_2}]^{\rm T}\ma B(\ve x_0)\ve n^{(0)}_{t_2}\left([\ve n^{\perp,(0)}_t]^{\rm T}\ma A(\ve x_0)\ve n^{(0)}_{t_1}+[\ve n^{(0)}_t]^{\rm T}\ma A(\ve x_0)\ve n^{\perp,(0)}_{t_1}\right)
\\&
\hspace{2.5cm}
+2[\ve n^{\perp,(0)}_{t_1}]^{\rm T}\ma B(\ve x_0)\ve n^{(0)}_{t_1}[\ve n^{\perp,(0)}_t]^{\rm T}\ma A(\ve x_0)\ve n^{(0)}_{t_2}
+2[\ve n^{(0)}_t]^{\rm T}\ma A(\ve x_0)\ma A(\ve x_0,t_1)\ve n^{(0)}_{t_2}
\Big\rangle\,.
\end{split}
\label{eq:Un_OddPhi}
\end{align}

Next we average over initial angles $\varphi_0$ and positions $\ve x_0$ using their distribution $P(\varphi_0,\ve x_0)$.
To zeroth order in $\ku$ there is no preferential concentration and the distribution is uniform, $P_0(\varphi_0,\ve x_0)=\ell_{\rm f}^2/(8\pi^3L^2)$, where $2\pi L/\ell_{\rm f}$ is the dimensionless side length of the domain.
Since $P_0$ is independent of $\ve x$, we use partial integration to simplify the terms proportional to $\ku^2$ in Eq.~(\ref{eq:Un_OddPhi}).
We assume that the mean flow $\ve U$ is either periodic or zero over the domain boundaries, this is the case for the Kolmogorov flow, so that the boundary terms vanish upon partial integrations.
Finally, we use that the stochastic part of the flow is statistically independent of the initial flow position $\ve x_0$ in the statistical model.
We obtain
\begin{align}
\begin{split}
&\langle\ve U(\ve x_t)\cdot\ve n_t\rangle
=
\int_0^{2\pi}{\rm d}\varphi_0\int{\rm d}^2\ve x_0\bigg[
-2\ku^2\Phi\frac{\ell_{\rm f}^2}{8\pi^3 L^2}\int_0^t{\rm d}t_1 t_1\big\langle\cos(\varphi^{(0)}_{t}+\varphi^{(0)}_{t_1})S_{12}(\ve x_0)O_{12}(\ve x_0)\big\rangle
\\&
+\ku\Phi\int_0^t{\rm d}t_1P(\varphi_0,\ve x_0)\langle S_{11}(\ve x_0)\cos(\varphi^{(0)}_t+\varphi^{(0)}_{t_1}) + S_{12}(\ve x_0)\sin(\varphi^{(0)}_t+\varphi^{(0)}_{t_1}) - O_{12}(\ve x_0)\sin(\varphi^{(0)}_t-\varphi^{(0)}_{t_1})
\rangle
\\&
+\ku^2\Phi\Lambda\frac{\ell_{\rm f}^2}{8\pi^3 L^2}\int_0^t{\rm d}t_1\int_0^{t_1}{\rm d}t_2\Big\langle
\cos(\varphi^{(0)}_{t}+\varphi^{(0)}_{t_1}+2\varphi^{(0)}_{t_2})(S_{12}(\ve x_0)^2-S_{11}(\ve x_0)^2)
\\&
\hspace{6.4cm}
+\cos(\varphi^{(0)}_{t} + \varphi^{(0)}_{t_1} - 2\varphi^{(0)}_{t_2})(S_{12}(\ve x_0)^2+S_{11}(\ve x_0)^2)\Big\rangle
\bigg]\,.
\end{split}
\label{eq:Un_x0AvgNew}
\end{align}

Next, we approximate $\varphi^{(0)}_t$, being the solution to Eq.~(\ref{eq:phi0Dot}), to first order in $\ku$
\begin{align}
\varphi^{(0)}_t&=\varphi_0-\ku\int_0^t{\rm d}t_1 \ve n^\perp_{0}\ma B(t_1)\ve n_0\,.
\end{align}
Within this approximation $\varphi^{(0)}_t$ is Gaussian distributed with correlation function
\begin{align}
\langle\varphi^{(0)}_{t}\varphi^{(0)}_{t'}\rangle
=\varphi_0^2+\ku^2\int_0^t{\rm d}t_1\int_0^{t'}{\rm d}t_2\Big\langle O_{12}(t_1)O_{12}(t_2)+\frac{1}{4}\Lambda^2{\rm Tr}(\ma S(t_1)\ma S(t_2))\Big\rangle\,.
\end{align}
For large times, the correlation function of the angles grow as $\sim {\rm min(t,t')}$ as expected from the diffusive nature of the angles when $\alpha=\Phi=0$.
Using that a sum over Gaussian variables is Gaussian distributed and that $\langle\cos(X)\rangle=\exp[-\tfrac{1}{2}\langle X^2\rangle]$ and $\langle\sin(X)\rangle=0$ for a Gaussian variable $X$, we conclude that in the steady state limit $t\to\infty$, the averages of $\cos(\varphi^{(0)}_{t}+\varphi^{(0)}_{t_1})$, $\cos(\varphi^{(0)}_{t}+\varphi^{(0)}_{t_1}+2\varphi^{(0)}_{t_2})$, $\sin(\varphi^{(0)}_{t}+\varphi^{(0)}_{t_1})$, and $\sin(\varphi^{(0)}_{t}-\varphi^{(0)}_{t_1})$ in Eq.~(\ref{eq:Un_x0AvgNew}) must vanish.
Averaging the remaining term in Eq.~(\ref{eq:Un_x0AvgNew}) for large $t$, the alignment becomes
\begin{align}
\label{eq:appendix_Un_General_result}
\langle\ve U\cdot\ve n_t\rangle&=
\frac{1}{2}\ku^2\Phi\Lambda\frac{\ell_{\rm f}^2}{(2\pi L)^2}\int{\rm d}^2\ve x_0{\rm Tr}(\ma S(\ve x_0)^2)\int_0^t{\rm d}t_1\int_0^{t_1}{\rm d}t_2
\\&\times
\exp\left[
-\frac{1}{2}\ku^2\left[
\int_{t_2}^t{\rm d}t'\int_{t_2}^t{\rm d}t''
+\int_{t_2}^{t_1}{\rm d}t'\int_{t_2}^{t_1}{\rm d}t''
+2\int_{t_2}^t{\rm d}t'\int_{t_2}^{t_1}{\rm d}t''
\right]
\langle O_{12}(t')O_{12}(t'')+\frac{1}{4}\Lambda^2{\rm Tr}(\ma S(t')\ma S(t''))\rangle
\right]\nonumber
\end{align}
This expression is identical to Eq.~(\ref{eq:projection_U}) in the main text, after using isotropy of the stochastic fluctuations to rewrite the correlation functions of $O_{12}$ and $\ma S$ in terms of the correlation function of $\ma A$.

We evaluate Eq.~(\ref{eq:appendix_Un_General_result}) explicitly by using the correlation functions for the statistical model
\begin{align}
\langle O_{12}(t)O_{12}(t')\rangle&=e^{-|t-t'|}\\
\langle S_{12}(t)S_{12}(t')\rangle&=\langle S_{11}(t)S_{11}(t')\rangle=\frac{1}{4}{\rm Tr}\langle\ma S(t)\ma S(t')\rangle=\frac{1}{2}e^{-|t-t'|}
\end{align}
and that $S_{11}(\ve x_0)=0$ and $S_{12}(\ve x_0)=-\kappa\alpha\sin(\kappa z_0)/2$ for the Kolmogorov flow to obtain
\begin{align}
\langle\ve U(\ve x_t)\cdot\ve n_t\rangle&=
\frac{1}{8}\ku^2\kappa^2\alpha^2\Phi\Lambda\int_0^t{\rm d}t_1\int_0^{t_1}{\rm d}t_2
\exp\left[
\ku^2\left(1+\tfrac{1}{2}\Lambda^2\right)\left[
3 + e^{t_1-t} - 2e^{t_2-t} - 2e^{t_2-t_1} - t - 3t_1 + 4t_2
\right]
\right]
\end{align}
Changing integration variables $t_1=t+s_2-s_1$ and $t_2=t-s_1$ and taking the limit $t\to\infty$ gives
\begin{align}
\langle\ve U(\ve x_t)\cdot\ve n_t\rangle&=
\frac{1}{8}\ku^2\kappa^2\alpha^2\Phi\Lambda\int_0^\infty{\rm d}t_1\int_0^{t_1}{\rm d}t_2
\exp\left[
\ku^2\left(1+\tfrac{1}{2}\Lambda^2\right)\left[
3 - 2e^{-t_2} - 2e^{-t_1} + e^{t_2-t_1} - 3 t_2 - t_1
\right]
\right]
\end{align}
In the limit of small values of $\ku$, we evaluate Eq.~(\ref{eq:projection_U_smallKu}) by replacing the exponential function by its large-time asymptote $\exp[\ku^2(1+\tfrac{1}{2}\Lambda^2)(- 3 t_2 - t_1)]$ before performing the integral. Expanding the resulting expression to lowest order in $\ku$ gives Eq.~(\ref{eq:projection_U_smallKu}) in the main text.

\section{Limit of large swimming speed}
\label{sec:large_Phi}

Starting from Eqs.~(\ref{eq:zz}) and (\ref{eq:phi}), the steady-state joint distribution $\rho(z,\varphi)$ of $z$ and $\varphi$ satisfies the continuity equation
\begin{align}
\frac{\partial\rho}{\partial t}&=-\frac{\partial}{\partial z}[\dot z\rho]-\frac{\partial}{\partial\varphi}[\dot\varphi\rho]
=-\Phi \sin\varphi\frac{\partial\rho}{\partial z}-\frac{1}{2}\sin z\frac{\partial}{\partial\varphi}[(1-\Lambda\cos(2\varphi))\rho]=0
\end{align}
Here lengths, velocities and times are dedimensionalized using the units $L$, $U_0$ and $L/U_0$ of Section~\ref{sec3}.
When $\Phi\to\infty$, we expect swimmers to be uniformly distributed over the $2\pi$-periodic coordinates $z$ and $\varphi$.
Making a series expansion to first order in $\Phi^{-1}$ around this limit in the continuity equation, gives the following distribution for large values of $\Phi$
\begin{align}
\rho(z,\varphi)\sim \frac{1}{4\pi^2}+\frac{1}{\Phi}\left[\frac{\Lambda}{2\pi^2}\cos z\cos\varphi+f(\varphi)\right]\,.
\end{align}
Here, following the symmetries in the Kolmogorov flow, $f(\varphi)$ is an undetermined symmetric $2\pi$-periodic function.
Averaging $\ve U\cdot\ve n=\cos z\cos\varphi$ using this distribution gives
\begin{align}
\langle\ve U\cdot\ve n\rangle=\frac{\Lambda}{2\Phi}\,,
\label{eq:projection_U_LargePhi_Appendix}
\end{align}
which is identical to Eq.~(\ref{eq:projection_U_LargePhi}) after changing to the units used there.

Equation~(\ref{eq:projection_U_LargePhi_Appendix}) agrees with the large $\Phi$ asymptote of $\langle\ve U\cdot\ve n\rangle$ evaluated using the deterministic dynamics in Eqs.~(\ref{eq:zz}) and (\ref{eq:phi}) [not shown].
Numerical simulations show that adding a small noise to the dynamics does not change this asymptotic behavior.
Moreover, the flow sampled along particle trajectories becomes white in time in the statistical model if the swimming speed is large enough, $v_{\rm s}\gg u_{\rm f}$, irrespective of the value of $\ku$.
As a consequence, the asymptote (\ref{eq:projection_U_LargePhi_Appendix}) holds for any value of $\ku$ if $v_{\rm s}$ is much larger than both $U$ and $u_{\rm f}$.

\bibliography{biblio}

\begin{thebibliography}{34}%
\makeatletter
\providecommand \@ifxundefined [1]{%
 \@ifx{#1\undefined}
}%
\providecommand \@ifnum [1]{%
 \ifnum #1\expandafter \@firstoftwo
 \else \expandafter \@secondoftwo
 \fi
}%
\providecommand \@ifx [1]{%
 \ifx #1\expandafter \@firstoftwo
 \else \expandafter \@secondoftwo
 \fi
}%
\providecommand \natexlab [1]{#1}%
\providecommand \enquote  [1]{``#1''}%
\providecommand \bibnamefont  [1]{#1}%
\providecommand \bibfnamefont [1]{#1}%
\providecommand \citenamefont [1]{#1}%
\providecommand \href@noop [0]{\@secondoftwo}%
\providecommand \href [0]{\begingroup \@sanitize@url \@href}%
\providecommand \@href[1]{\@@startlink{#1}\@@href}%
\providecommand \@@href[1]{\endgroup#1\@@endlink}%
\providecommand \@sanitize@url [0]{\catcode `\\12\catcode `\$12\catcode
  `\&12\catcode `\#12\catcode `\^12\catcode `\_12\catcode `\%12\relax}%
\providecommand \@@startlink[1]{}%
\providecommand \@@endlink[0]{}%
\providecommand \url  [0]{\begingroup\@sanitize@url \@url }%
\providecommand \@url [1]{\endgroup\@href {#1}{\urlprefix }}%
\providecommand \urlprefix  [0]{URL }%
\providecommand \Eprint [0]{\href }%
\providecommand \doibase [0]{https://doi.org/}%
\providecommand \selectlanguage [0]{\@gobble}%
\providecommand \bibinfo  [0]{\@secondoftwo}%
\providecommand \bibfield  [0]{\@secondoftwo}%
\providecommand \translation [1]{[#1]}%
\providecommand \BibitemOpen [0]{}%
\providecommand \bibitemStop [0]{}%
\providecommand \bibitemNoStop [0]{.\EOS\space}%
\providecommand \EOS [0]{\spacefactor3000\relax}%
\providecommand \BibitemShut  [1]{\csname bibitem#1\endcsname}%
\let\auto@bib@innerbib\@empty
\bibitem [{\citenamefont {Guasto}\ \emph {et~al.}(2012)\citenamefont {Guasto},
  \citenamefont {Rusconi},\ and\ \citenamefont {Stocker}}]{guasto2012}%
  \BibitemOpen
  \bibfield  {author} {\bibinfo {author} {\bibfnamefont {J.~S.}\ \bibnamefont
  {Guasto}}, \bibinfo {author} {\bibfnamefont {R.}~\bibnamefont {Rusconi}},\
  and\ \bibinfo {author} {\bibfnamefont {R.}~\bibnamefont {Stocker}},\
  }\bibfield  {title} {\bibinfo {title} {Fluid mechanics of planktonic
  microorganisms},\ }\href@noop {} {\bibfield  {journal} {\bibinfo  {journal}
  {Annu. Rev. Fluid Mech.}\ }\textbf {\bibinfo {volume} {44}},\ \bibinfo
  {pages} {373} (\bibinfo {year} {2012})}\BibitemShut {NoStop}%
\bibitem [{\citenamefont {Ki{\o}rboe}(2008)}]{kiorboe2008}%
  \BibitemOpen
  \bibfield  {author} {\bibinfo {author} {\bibfnamefont {T.}~\bibnamefont
  {Ki{\o}rboe}},\ }\href@noop {} {\emph {\bibinfo {title} {A mechanistic
  approach to plankton ecology}}}\ (\bibinfo  {publisher} {Princeton University
  Press},\ \bibinfo {year} {2008})\BibitemShut {NoStop}%
\bibitem [{\citenamefont {Z{\"o}ttl}\ and\ \citenamefont
  {Stark}(2012)}]{zottl2012nonlinear}%
  \BibitemOpen
  \bibfield  {author} {\bibinfo {author} {\bibfnamefont {A.}~\bibnamefont
  {Z{\"o}ttl}}\ and\ \bibinfo {author} {\bibfnamefont {H.}~\bibnamefont
  {Stark}},\ }\bibfield  {title} {\bibinfo {title} {Nonlinear dynamics of a
  microswimmer in {P}oiseuille flow},\ }\href@noop {} {\bibfield  {journal}
  {\bibinfo  {journal} {Phys. Rev. Lett.}\ }\textbf {\bibinfo {volume} {108}},\
  \bibinfo {pages} {218104} (\bibinfo {year} {2012})}\BibitemShut {NoStop}%
\bibitem [{\citenamefont {Z{\"o}ttl}\ and\ \citenamefont
  {Stark}(2013)}]{zottl2013periodic}%
  \BibitemOpen
  \bibfield  {author} {\bibinfo {author} {\bibfnamefont {A.}~\bibnamefont
  {Z{\"o}ttl}}\ and\ \bibinfo {author} {\bibfnamefont {H.}~\bibnamefont
  {Stark}},\ }\bibfield  {title} {\bibinfo {title} {Periodic and quasiperiodic
  motion of an elongated microswimmer in {P}oiseuille flow},\ }\href@noop {}
  {\bibfield  {journal} {\bibinfo  {journal} {Eur. Phys. J. E}\ }\textbf
  {\bibinfo {volume} {36}},\ \bibinfo {pages} {1} (\bibinfo {year}
  {2013})}\BibitemShut {NoStop}%
\bibitem [{\citenamefont {Junot}\ \emph {et~al.}(2019)\citenamefont {Junot},
  \citenamefont {Figueroa-Morales}, \citenamefont {Darnige}, \citenamefont
  {Lindner}, \citenamefont {Soto}, \citenamefont {Auradou},\ and\ \citenamefont
  {Cl{\'e}ment}}]{junot2019swimming}%
  \BibitemOpen
  \bibfield  {author} {\bibinfo {author} {\bibfnamefont {G.}~\bibnamefont
  {Junot}}, \bibinfo {author} {\bibfnamefont {N.}~\bibnamefont
  {Figueroa-Morales}}, \bibinfo {author} {\bibfnamefont {T.}~\bibnamefont
  {Darnige}}, \bibinfo {author} {\bibfnamefont {A.}~\bibnamefont {Lindner}},
  \bibinfo {author} {\bibfnamefont {R.}~\bibnamefont {Soto}}, \bibinfo {author}
  {\bibfnamefont {H.}~\bibnamefont {Auradou}},\ and\ \bibinfo {author}
  {\bibfnamefont {E.}~\bibnamefont {Cl{\'e}ment}},\ }\bibfield  {title}
  {\bibinfo {title} {Swimming bacteria in {P}oiseuille flow: The quest for
  active {B}retherton-{J}effery trajectories},\ }\href@noop {} {\bibfield
  {journal} {\bibinfo  {journal} {EPL}\ }\textbf {\bibinfo {volume} {126}},\
  \bibinfo {pages} {44003} (\bibinfo {year} {2019})}\BibitemShut {NoStop}%
\bibitem [{\citenamefont {Berman}\ \emph {et~al.}(2022)\citenamefont {Berman},
  \citenamefont {Ferguson}, \citenamefont {Bizzak}, \citenamefont {Solomon},\
  and\ \citenamefont {Mitchell}}]{berman2021swimmer}%
  \BibitemOpen
  \bibfield  {author} {\bibinfo {author} {\bibfnamefont {S.~A.}\ \bibnamefont
  {Berman}}, \bibinfo {author} {\bibfnamefont {K.~S.}\ \bibnamefont
  {Ferguson}}, \bibinfo {author} {\bibfnamefont {N.}~\bibnamefont {Bizzak}},
  \bibinfo {author} {\bibfnamefont {T.~H.}\ \bibnamefont {Solomon}},\ and\
  \bibinfo {author} {\bibfnamefont {K.~A.}\ \bibnamefont {Mitchell}},\
  }\bibfield  {title} {\bibinfo {title} {Noise-induced aggregation of swimmers
  in the {K}olmogorov flow},\ }\href@noop {} {\bibfield  {journal} {\bibinfo
  {journal} {Front. Phys.}\ }\textbf {\bibinfo {volume} {9}},\ \bibinfo {pages}
  {816663} (\bibinfo {year} {2022})}\BibitemShut {NoStop}%
\bibitem [{\citenamefont {Rusconi}\ \emph {et~al.}(2014)\citenamefont
  {Rusconi}, \citenamefont {Guasto},\ and\ \citenamefont
  {Stocker}}]{rusconi2014bacterial}%
  \BibitemOpen
  \bibfield  {author} {\bibinfo {author} {\bibfnamefont {R.}~\bibnamefont
  {Rusconi}}, \bibinfo {author} {\bibfnamefont {J.~S.}\ \bibnamefont
  {Guasto}},\ and\ \bibinfo {author} {\bibfnamefont {R.}~\bibnamefont
  {Stocker}},\ }\bibfield  {title} {\bibinfo {title} {Bacterial transport
  suppressed by fluid shear},\ }\href@noop {} {\bibfield  {journal} {\bibinfo
  {journal} {Nat. Phys.}\ }\textbf {\bibinfo {volume} {10}},\ \bibinfo {pages}
  {212} (\bibinfo {year} {2014})}\BibitemShut {NoStop}%
\bibitem [{\citenamefont {Bearon}\ and\ \citenamefont
  {Hazel}(2015)}]{bearon2015trapping}%
  \BibitemOpen
  \bibfield  {author} {\bibinfo {author} {\bibfnamefont {R.}~\bibnamefont
  {Bearon}}\ and\ \bibinfo {author} {\bibfnamefont {A.}~\bibnamefont {Hazel}},\
  }\bibfield  {title} {\bibinfo {title} {The trapping in high-shear regions of
  slender bacteria undergoing chemotaxis in a channel},\ }\href@noop {}
  {\bibfield  {journal} {\bibinfo  {journal} {J. Fluid Mech.}\ }\textbf
  {\bibinfo {volume} {771}} (\bibinfo {year} {2015})}\BibitemShut {NoStop}%
\bibitem [{\citenamefont {Durham}\ \emph {et~al.}(2009)\citenamefont {Durham},
  \citenamefont {Kessler},\ and\ \citenamefont
  {Stocker}}]{durham2009disruption}%
  \BibitemOpen
  \bibfield  {author} {\bibinfo {author} {\bibfnamefont {W.~M.}\ \bibnamefont
  {Durham}}, \bibinfo {author} {\bibfnamefont {J.~O.}\ \bibnamefont
  {Kessler}},\ and\ \bibinfo {author} {\bibfnamefont {R.}~\bibnamefont
  {Stocker}},\ }\bibfield  {title} {\bibinfo {title} {Disruption of vertical
  motility by shear triggers formation of thin phytoplankton layers},\
  }\href@noop {} {\bibfield  {journal} {\bibinfo  {journal} {Science}\ }\textbf
  {\bibinfo {volume} {323}},\ \bibinfo {pages} {1067} (\bibinfo {year}
  {2009})}\BibitemShut {NoStop}%
\bibitem [{\citenamefont {Santamaria}\ \emph {et~al.}(2014)\citenamefont
  {Santamaria}, \citenamefont {De~Lillo}, \citenamefont {Cencini},\ and\
  \citenamefont {Boffetta}}]{santamaria2014gyrotactic}%
  \BibitemOpen
  \bibfield  {author} {\bibinfo {author} {\bibfnamefont {F.}~\bibnamefont
  {Santamaria}}, \bibinfo {author} {\bibfnamefont {F.}~\bibnamefont
  {De~Lillo}}, \bibinfo {author} {\bibfnamefont {M.}~\bibnamefont {Cencini}},\
  and\ \bibinfo {author} {\bibfnamefont {G.}~\bibnamefont {Boffetta}},\
  }\bibfield  {title} {\bibinfo {title} {Gyrotactic trapping in laminar and
  turbulent {K}olmogorov flow},\ }\href@noop {} {\bibfield  {journal} {\bibinfo
   {journal} {Phys. Fluids}\ }\textbf {\bibinfo {volume} {26}},\ \bibinfo
  {pages} {111901} (\bibinfo {year} {2014})}\BibitemShut {NoStop}%
\bibitem [{\citenamefont {Barry}\ \emph {et~al.}(2015)\citenamefont {Barry},
  \citenamefont {Rusconi}, \citenamefont {Guasto},\ and\ \citenamefont
  {Stocker}}]{barry2015shear}%
  \BibitemOpen
  \bibfield  {author} {\bibinfo {author} {\bibfnamefont {M.~T.}\ \bibnamefont
  {Barry}}, \bibinfo {author} {\bibfnamefont {R.}~\bibnamefont {Rusconi}},
  \bibinfo {author} {\bibfnamefont {J.~S.}\ \bibnamefont {Guasto}},\ and\
  \bibinfo {author} {\bibfnamefont {R.}~\bibnamefont {Stocker}},\ }\bibfield
  {title} {\bibinfo {title} {Shear-induced orientational dynamics and spatial
  heterogeneity in suspensions of motile phytoplankton},\ }\href@noop {}
  {\bibfield  {journal} {\bibinfo  {journal} {J. Royal Soc. Interface}\
  }\textbf {\bibinfo {volume} {12}},\ \bibinfo {pages} {20150791} (\bibinfo
  {year} {2015})}\BibitemShut {NoStop}%
\bibitem [{\citenamefont {Arguedas-Leiva}\ and\ \citenamefont
  {Wilczek}(2020)}]{arguedas2020microswimmers}%
  \BibitemOpen
  \bibfield  {author} {\bibinfo {author} {\bibfnamefont {J.-A.}\ \bibnamefont
  {Arguedas-Leiva}}\ and\ \bibinfo {author} {\bibfnamefont {M.}~\bibnamefont
  {Wilczek}},\ }\bibfield  {title} {\bibinfo {title} {Microswimmers in an
  axisymmetric vortex flow},\ }\href@noop {} {\bibfield  {journal} {\bibinfo
  {journal} {New J. Phys.}\ }\textbf {\bibinfo {volume} {22}},\ \bibinfo
  {pages} {053051} (\bibinfo {year} {2020})}\BibitemShut {NoStop}%
\bibitem [{\citenamefont {Dehkharghani}\ \emph {et~al.}(2019)\citenamefont
  {Dehkharghani}, \citenamefont {Waisbord}, \citenamefont {Dunkel},\ and\
  \citenamefont {Guasto}}]{dehkharghani2019bacterial}%
  \BibitemOpen
  \bibfield  {author} {\bibinfo {author} {\bibfnamefont {A.}~\bibnamefont
  {Dehkharghani}}, \bibinfo {author} {\bibfnamefont {N.}~\bibnamefont
  {Waisbord}}, \bibinfo {author} {\bibfnamefont {J.}~\bibnamefont {Dunkel}},\
  and\ \bibinfo {author} {\bibfnamefont {J.~S.}\ \bibnamefont {Guasto}},\
  }\bibfield  {title} {\bibinfo {title} {Bacterial scattering in microfluidic
  crystal flows reveals giant active {T}aylor--{A}ris dispersion},\ }\href@noop
  {} {\bibfield  {journal} {\bibinfo  {journal} {Proc. Natl. Acad. Sci.
  U.S.A.}\ }\textbf {\bibinfo {volume} {116}},\ \bibinfo {pages} {11119}
  (\bibinfo {year} {2019})}\BibitemShut {NoStop}%
\bibitem [{\citenamefont {Zhan}\ \emph {et~al.}(2014)\citenamefont {Zhan},
  \citenamefont {Sardina}, \citenamefont {Lushi},\ and\ \citenamefont
  {Brandt}}]{zhan2014accumulation}%
  \BibitemOpen
  \bibfield  {author} {\bibinfo {author} {\bibfnamefont {C.}~\bibnamefont
  {Zhan}}, \bibinfo {author} {\bibfnamefont {G.}~\bibnamefont {Sardina}},
  \bibinfo {author} {\bibfnamefont {E.}~\bibnamefont {Lushi}},\ and\ \bibinfo
  {author} {\bibfnamefont {L.}~\bibnamefont {Brandt}},\ }\bibfield  {title}
  {\bibinfo {title} {Accumulation of motile elongated micro-organisms in
  turbulence},\ }\href@noop {} {\bibfield  {journal} {\bibinfo  {journal} {J.
  Fluid Mech.}\ }\textbf {\bibinfo {volume} {739}},\ \bibinfo {pages} {22}
  (\bibinfo {year} {2014})}\BibitemShut {NoStop}%
\bibitem [{\citenamefont {Pujara}\ \emph {et~al.}(2018)\citenamefont {Pujara},
  \citenamefont {Koehl},\ and\ \citenamefont {Variano}}]{pujara2018rotations}%
  \BibitemOpen
  \bibfield  {author} {\bibinfo {author} {\bibfnamefont {N.}~\bibnamefont
  {Pujara}}, \bibinfo {author} {\bibfnamefont {M.}~\bibnamefont {Koehl}},\ and\
  \bibinfo {author} {\bibfnamefont {E.}~\bibnamefont {Variano}},\ }\bibfield
  {title} {\bibinfo {title} {Rotations and accumulation of ellipsoidal
  microswimmers in isotropic turbulence},\ }\href@noop {} {\bibfield  {journal}
  {\bibinfo  {journal} {J. Fluid Mech.}\ }\textbf {\bibinfo {volume} {838}},\
  \bibinfo {pages} {356} (\bibinfo {year} {2018})}\BibitemShut {NoStop}%
\bibitem [{\citenamefont {Yazdi}\ and\ \citenamefont
  {Ardekani}(2012)}]{yazdi2012bacterial}%
  \BibitemOpen
  \bibfield  {author} {\bibinfo {author} {\bibfnamefont {S.}~\bibnamefont
  {Yazdi}}\ and\ \bibinfo {author} {\bibfnamefont {A.~M.}\ \bibnamefont
  {Ardekani}},\ }\bibfield  {title} {\bibinfo {title} {Bacterial aggregation
  and biofilm formation in a vortical flow},\ }\href
  {https://doi.org/10.1063/1.4771407} {\bibfield  {journal} {\bibinfo
  {journal} {Biomicrofluidics}\ }\textbf {\bibinfo {volume} {6}},\ \bibinfo
  {pages} {044114} (\bibinfo {year} {2012})}\BibitemShut {NoStop}%
\bibitem [{\citenamefont {Borgnino}\ \emph {et~al.}(2019)\citenamefont
  {Borgnino}, \citenamefont {Gustavsson}, \citenamefont {De~Lillo},
  \citenamefont {Boffetta}, \citenamefont {Cencini},\ and\ \citenamefont
  {Mehlig}}]{borgnino2019alignment}%
  \BibitemOpen
  \bibfield  {author} {\bibinfo {author} {\bibfnamefont {M.}~\bibnamefont
  {Borgnino}}, \bibinfo {author} {\bibfnamefont {K.}~\bibnamefont
  {Gustavsson}}, \bibinfo {author} {\bibfnamefont {F.}~\bibnamefont
  {De~Lillo}}, \bibinfo {author} {\bibfnamefont {G.}~\bibnamefont {Boffetta}},
  \bibinfo {author} {\bibfnamefont {M.}~\bibnamefont {Cencini}},\ and\ \bibinfo
  {author} {\bibfnamefont {B.}~\bibnamefont {Mehlig}},\ }\bibfield  {title}
  {\bibinfo {title} {Alignment of nonspherical active particles in chaotic
  flows},\ }\href@noop {} {\bibfield  {journal} {\bibinfo  {journal} {Phys.
  Rev. Lett.}\ }\textbf {\bibinfo {volume} {123}},\ \bibinfo {pages} {138003}
  (\bibinfo {year} {2019})}\BibitemShut {NoStop}%
\bibitem [{\citenamefont {Meshalkin}\ and\ \citenamefont
  {Sinai}(1961)}]{meshalkin1961investigation}%
  \BibitemOpen
  \bibfield  {author} {\bibinfo {author} {\bibfnamefont {L.}~\bibnamefont
  {Meshalkin}}\ and\ \bibinfo {author} {\bibfnamefont {I.~G.}\ \bibnamefont
  {Sinai}},\ }\bibfield  {title} {\bibinfo {title} {Investigation of the
  stability of a stationary solution of a system of equations for the plane
  movement of an incompressible viscous liquid},\ }\href@noop {} {\bibfield
  {journal} {\bibinfo  {journal} {J. Appl. Math. Mech.}\ }\textbf {\bibinfo
  {volume} {25}},\ \bibinfo {pages} {1700} (\bibinfo {year}
  {1961})}\BibitemShut {NoStop}%
\bibitem [{\citenamefont {Musacchio}\ and\ \citenamefont
  {Boffetta}(2014)}]{musacchio2014turbulent}%
  \BibitemOpen
  \bibfield  {author} {\bibinfo {author} {\bibfnamefont {S.}~\bibnamefont
  {Musacchio}}\ and\ \bibinfo {author} {\bibfnamefont {G.}~\bibnamefont
  {Boffetta}},\ }\bibfield  {title} {\bibinfo {title} {Turbulent channel
  without boundaries: The periodic {K}olmogorov flow},\ }\href@noop {}
  {\bibfield  {journal} {\bibinfo  {journal} {Phys. Rev. E}\ }\textbf {\bibinfo
  {volume} {89}},\ \bibinfo {pages} {023004} (\bibinfo {year}
  {2014})}\BibitemShut {NoStop}%
\bibitem [{\citenamefont {Gustavsson}\ and\ \citenamefont
  {Mehlig}(2016)}]{Gustavsson2016}%
  \BibitemOpen
  \bibfield  {author} {\bibinfo {author} {\bibfnamefont {K.}~\bibnamefont
  {Gustavsson}}\ and\ \bibinfo {author} {\bibfnamefont {B.}~\bibnamefont
  {Mehlig}},\ }\bibfield  {title} {\bibinfo {title} {Statistical models for
  spatial patterns of heavy particles in turbulence},\ }\href
  {https://doi.org/10.1080/00018732.2016.1164490} {\bibfield  {journal}
  {\bibinfo  {journal} {Adv. Phys.}\ }\textbf {\bibinfo {volume} {65}},\
  \bibinfo {pages} {1} (\bibinfo {year} {2016})}\BibitemShut {NoStop}%
\bibitem [{\citenamefont {Pedley}\ and\ \citenamefont
  {Kessler}(1987)}]{pedley87}%
  \BibitemOpen
  \bibfield  {author} {\bibinfo {author} {\bibfnamefont {T.~J.}\ \bibnamefont
  {Pedley}}\ and\ \bibinfo {author} {\bibfnamefont {J.~O.}\ \bibnamefont
  {Kessler}},\ }\bibfield  {title} {\bibinfo {title} {The orientation of
  spheroidal microorganisms swimming in a flow field},\ }\href
  {https://doi.org/10.1098/rspb.1987.0035} {\bibfield  {journal} {\bibinfo
  {journal} {Proc. R. Soc. Lond.}\ }\textbf {\bibinfo {volume} {231}},\
  \bibinfo {pages} {47} (\bibinfo {year} {1987})}\BibitemShut {NoStop}%
\bibitem [{\citenamefont {Jeffery}(1922)}]{jeffery1922}%
  \BibitemOpen
  \bibfield  {author} {\bibinfo {author} {\bibfnamefont {G.~B.}\ \bibnamefont
  {Jeffery}},\ }\bibfield  {title} {\bibinfo {title} {The motion of ellipsoidal
  particles immersed in a viscous fluid},\ }\href
  {https://doi.org/10.1098/rspa.1922.0078} {\bibfield  {journal} {\bibinfo
  {journal} {Proc. R. Soc. Lond.}\ }\textbf {\bibinfo {volume} {102}},\
  \bibinfo {pages} {161} (\bibinfo {year} {1922})}\BibitemShut {NoStop}%
\bibitem [{\citenamefont {Frisch}(1995)}]{frisch1995turbulence}%
  \BibitemOpen
  \bibfield  {author} {\bibinfo {author} {\bibfnamefont {U.}~\bibnamefont
  {Frisch}},\ }\href@noop {} {\emph {\bibinfo {title} {Turbulence: the legacy
  of {A}.{N}. {K}olmogorov}}}\ (\bibinfo  {publisher} {Cambridge University
  Press},\ \bibinfo {year} {1995})\BibitemShut {NoStop}%
\bibitem [{\citenamefont {Ardekani}\ \emph {et~al.}(2017)\citenamefont
  {Ardekani}, \citenamefont {Sardina}, \citenamefont {Brandt}, \citenamefont
  {Karp-Boss}, \citenamefont {Bearon},\ and\ \citenamefont
  {Variano}}]{ardekani2017sedimentation}%
  \BibitemOpen
  \bibfield  {author} {\bibinfo {author} {\bibfnamefont {M.~N.}\ \bibnamefont
  {Ardekani}}, \bibinfo {author} {\bibfnamefont {G.}~\bibnamefont {Sardina}},
  \bibinfo {author} {\bibfnamefont {L.}~\bibnamefont {Brandt}}, \bibinfo
  {author} {\bibfnamefont {L.}~\bibnamefont {Karp-Boss}}, \bibinfo {author}
  {\bibfnamefont {R.~N.}\ \bibnamefont {Bearon}},\ and\ \bibinfo {author}
  {\bibfnamefont {E.~A.}\ \bibnamefont {Variano}},\ }\bibfield  {title}
  {\bibinfo {title} {Sedimentation of inertia-less prolate spheroids in
  homogenous isotropic turbulence with application to non-motile
  phytoplankton},\ }\href@noop {} {\bibfield  {journal} {\bibinfo  {journal}
  {J. Fluid Mech.}\ }\textbf {\bibinfo {volume} {831}},\ \bibinfo {pages} {655}
  (\bibinfo {year} {2017})}\BibitemShut {NoStop}%
\bibitem [{\citenamefont {Gustavsson}\ and\ \citenamefont
  {Mehlig}(2011)}]{Gustavsson2011Ergodic}%
  \BibitemOpen
  \bibfield  {author} {\bibinfo {author} {\bibfnamefont {K.}~\bibnamefont
  {Gustavsson}}\ and\ \bibinfo {author} {\bibfnamefont {B.}~\bibnamefont
  {Mehlig}},\ }\bibfield  {title} {\bibinfo {title} {Ergodic and non-ergodic
  clustering of inertial particles},\ }\href
  {https://doi.org/10.1209/0295-5075/96/60012} {\bibfield  {journal} {\bibinfo
  {journal} {EPL}\ }\textbf {\bibinfo {volume} {96}},\ \bibinfo {pages} {60012}
  (\bibinfo {year} {2011})}\BibitemShut {NoStop}%
\bibitem [{\citenamefont {Gustavsson}\ \emph {et~al.}(2014)\citenamefont
  {Gustavsson}, \citenamefont {Vajedi},\ and\ \citenamefont
  {Mehlig}}]{Gustavsson2014Clustering}%
  \BibitemOpen
  \bibfield  {author} {\bibinfo {author} {\bibfnamefont {K.}~\bibnamefont
  {Gustavsson}}, \bibinfo {author} {\bibfnamefont {S.}~\bibnamefont {Vajedi}},\
  and\ \bibinfo {author} {\bibfnamefont {B.}~\bibnamefont {Mehlig}},\
  }\bibfield  {title} {\bibinfo {title} {Clustering of particles falling in a
  turbulent flow},\ }\href {https://doi.org/10.1103/physrevlett.112.214501}
  {\bibfield  {journal} {\bibinfo  {journal} {Phys. Rev. Lett.}\ }\textbf
  {\bibinfo {volume} {112}},\ \bibinfo {pages} {214501} (\bibinfo {year}
  {2014})}\BibitemShut {NoStop}%
\bibitem [{\citenamefont {Gustavsson}\ \emph {et~al.}(2016)\citenamefont
  {Gustavsson}, \citenamefont {Berglund}, \citenamefont {Jonsson},\ and\
  \citenamefont {Mehlig}}]{gustavsson2016gyrotaxis}%
  \BibitemOpen
  \bibfield  {author} {\bibinfo {author} {\bibfnamefont {K.}~\bibnamefont
  {Gustavsson}}, \bibinfo {author} {\bibfnamefont {F.}~\bibnamefont
  {Berglund}}, \bibinfo {author} {\bibfnamefont {P.~R.}\ \bibnamefont
  {Jonsson}},\ and\ \bibinfo {author} {\bibfnamefont {B.}~\bibnamefont
  {Mehlig}},\ }\bibfield  {title} {\bibinfo {title} {Preferential sampling and
  small-scale clustering of gyrotactic microswimmers in turbulence},\
  }\href@noop {} {\bibfield  {journal} {\bibinfo  {journal} {Phys. Rev. Lett.}\
  }\textbf {\bibinfo {volume} {116}},\ \bibinfo {pages} {108104} (\bibinfo
  {year} {2016})}\BibitemShut {NoStop}%
\bibitem [{\citenamefont {Gustavsson}\ \emph {et~al.}(2017)\citenamefont
  {Gustavsson}, \citenamefont {Jucha}, \citenamefont {Naso}, \citenamefont
  {L{\'{e}}v{\^{e}}que}, \citenamefont {Pumir},\ and\ \citenamefont
  {Mehlig}}]{Gustavsson2017Statistical}%
  \BibitemOpen
  \bibfield  {author} {\bibinfo {author} {\bibfnamefont {K.}~\bibnamefont
  {Gustavsson}}, \bibinfo {author} {\bibfnamefont {J.}~\bibnamefont {Jucha}},
  \bibinfo {author} {\bibfnamefont {A.}~\bibnamefont {Naso}}, \bibinfo {author}
  {\bibfnamefont {E.}~\bibnamefont {L{\'{e}}v{\^{e}}que}}, \bibinfo {author}
  {\bibfnamefont {A.}~\bibnamefont {Pumir}},\ and\ \bibinfo {author}
  {\bibfnamefont {B.}~\bibnamefont {Mehlig}},\ }\bibfield  {title} {\bibinfo
  {title} {Statistical model for the orientation of nonspherical particles
  settling in turbulence},\ }\href
  {https://doi.org/10.1103/physrevlett.119.254501} {\bibfield  {journal}
  {\bibinfo  {journal} {Phys. Rev. Lett.}\ }\textbf {\bibinfo {volume} {119}},\
  \bibinfo {pages} {254501} (\bibinfo {year} {2017})}\BibitemShut {NoStop}%
\bibitem [{\citenamefont {Shroyer}\ \emph {et~al.}(2014)\citenamefont
  {Shroyer}, \citenamefont {Benoit-Bird}, \citenamefont {Nash},\ and\
  \citenamefont {Moum}}]{shroyer2014stratification}%
  \BibitemOpen
  \bibfield  {author} {\bibinfo {author} {\bibfnamefont {E.~L.}\ \bibnamefont
  {Shroyer}}, \bibinfo {author} {\bibfnamefont {K.~J.}\ \bibnamefont
  {Benoit-Bird}}, \bibinfo {author} {\bibfnamefont {J.~D.}\ \bibnamefont
  {Nash}},\ and\ \bibinfo {author} {\bibfnamefont {J.~N.}\ \bibnamefont
  {Moum}},\ }\bibfield  {title} {\bibinfo {title} {Stratification and mixing
  regimes in biological thin layers over the {Mid}-{Atlantic} {Bight}},\
  }\href@noop {} {\bibfield  {journal} {\bibinfo  {journal} {Limnol.
  Oceanogr.}\ }\textbf {\bibinfo {volume} {59}},\ \bibinfo {pages} {1349}
  (\bibinfo {year} {2014})}\BibitemShut {NoStop}%
\bibitem [{\citenamefont {Durham}\ and\ \citenamefont
  {Stocker}(2012)}]{durham2012thin}%
  \BibitemOpen
  \bibfield  {author} {\bibinfo {author} {\bibfnamefont {W.~M.}\ \bibnamefont
  {Durham}}\ and\ \bibinfo {author} {\bibfnamefont {R.}~\bibnamefont
  {Stocker}},\ }\bibfield  {title} {\bibinfo {title} {Thin phytoplankton
  layers: characteristics, mechanisms, and consequences},\ }\href@noop {}
  {\bibfield  {journal} {\bibinfo  {journal} {Annu. Rev. Mar. Sci.}\ }\textbf
  {\bibinfo {volume} {4}},\ \bibinfo {pages} {177} (\bibinfo {year}
  {2012})}\BibitemShut {NoStop}%
\bibitem [{\citenamefont {Franks}\ \emph {et~al.}(2022)\citenamefont {Franks},
  \citenamefont {Inman}, \citenamefont {MacKinnon}, \citenamefont {Alford},\
  and\ \citenamefont {Waterhouse}}]{franks2022oceanic}%
  \BibitemOpen
  \bibfield  {author} {\bibinfo {author} {\bibfnamefont {P.~J.}\ \bibnamefont
  {Franks}}, \bibinfo {author} {\bibfnamefont {B.~G.}\ \bibnamefont {Inman}},
  \bibinfo {author} {\bibfnamefont {J.~A.}\ \bibnamefont {MacKinnon}}, \bibinfo
  {author} {\bibfnamefont {M.~H.}\ \bibnamefont {Alford}},\ and\ \bibinfo
  {author} {\bibfnamefont {A.~F.}\ \bibnamefont {Waterhouse}},\ }\bibfield
  {title} {\bibinfo {title} {Oceanic turbulence from a planktonic
  perspective},\ }\href@noop {} {\bibfield  {journal} {\bibinfo  {journal}
  {Limnol. and Oceanogr.}\ }\textbf {\bibinfo {volume} {67}},\ \bibinfo {pages}
  {348} (\bibinfo {year} {2022})}\BibitemShut {NoStop}%
\bibitem [{\citenamefont {Thorpe}(2007)}]{thorpe2007introduction}%
  \BibitemOpen
  \bibfield  {author} {\bibinfo {author} {\bibfnamefont {S.~A.}\ \bibnamefont
  {Thorpe}},\ }\href@noop {} {\emph {\bibinfo {title} {{A}n {I}ntroduction to
  {O}cean {T}urbulence}}}\ (\bibinfo  {publisher} {Cambridge University
  Press},\ \bibinfo {year} {2007})\BibitemShut {NoStop}%
\bibitem [{\citenamefont {Yamazaki}\ and\ \citenamefont
  {Squires}(1996)}]{yamazaki1996comparison}%
  \BibitemOpen
  \bibfield  {author} {\bibinfo {author} {\bibfnamefont {H.}~\bibnamefont
  {Yamazaki}}\ and\ \bibinfo {author} {\bibfnamefont {K.~D.}\ \bibnamefont
  {Squires}},\ }\bibfield  {title} {\bibinfo {title} {Comparison of oceanic
  turbulence and copepod swimming},\ }\href@noop {} {\bibfield  {journal}
  {\bibinfo  {journal} {Mar. Ecol. Prog. Ser.}\ }\textbf {\bibinfo {volume}
  {144}},\ \bibinfo {pages} {299} (\bibinfo {year} {1996})}\BibitemShut
  {NoStop}%
\bibitem [{\citenamefont {Lovecchio}\ \emph {et~al.}(2019)\citenamefont
  {Lovecchio}, \citenamefont {Climent}, \citenamefont {Stocker},\ and\
  \citenamefont {Durham}}]{lovecchio2019chain}%
  \BibitemOpen
  \bibfield  {author} {\bibinfo {author} {\bibfnamefont {S.}~\bibnamefont
  {Lovecchio}}, \bibinfo {author} {\bibfnamefont {E.}~\bibnamefont {Climent}},
  \bibinfo {author} {\bibfnamefont {R.}~\bibnamefont {Stocker}},\ and\ \bibinfo
  {author} {\bibfnamefont {W.~M.}\ \bibnamefont {Durham}},\ }\bibfield  {title}
  {\bibinfo {title} {Chain formation can enhance the vertical migration of
  phytoplankton through turbulence},\ }\href@noop {} {\bibfield  {journal}
  {\bibinfo  {journal} {Sci. Adv.}\ }\textbf {\bibinfo {volume} {5}},\ \bibinfo
  {pages} {eaaw7879} (\bibinfo {year} {2019})}\BibitemShut {NoStop}%
\end{thebibliography}%

\end{document}